\documentclass{iopjournal}

\usepackage{amsmath,amsfonts,amssymb,bm} 
\usepackage{epsf,mathrsfs}

	\newcommand{\exgra}{{\em e.g.}}
	\newcommand{\idest}{{\em i.e.}}
	\newcommand{\sgn}{{\rm sgn}}

\begin{document}

\articletype{Paper} 

\title{On the kinetic continuous spectrum in toroidal plasmas}

\author{Fulvio Zonca$^{1,2,*}$\orcid{0000-0002-9270-4704}, Liu Chen$^{3,1}$\orcid{0000-0002-4180-588X}, Matteo V. Falessi$^{1,4}$\orcid{0000-0002-2105-226X}, 
Ruirui Ma$^{5,1}$\orcid{0000-0002-0169-0643}, Zhiyong Qiu$^{6,1}$\orcid{0000-0002-7548-8819}
 and Guangyu Wei$^{3}$\orcid{0000-0001-8555-4229}}

\affil{$^1$Center for Nonlinear Plasma Science and C.R. ENEA Frascati, C.P. 65, Frascati 00044, Italy}

\affil{$^2$Institute for Advanced Study in Physics, Zhejiang University, Hangzhou 310027, People’s Republic of China}

\affil{$^3$Institute for Fusion Theory and Simulation, School of Physics, Zhejiang University, Hangzhou 310027, People’s Republic of China}

\affil{$^4$Istituto Nazionale di Fisica Nucleare (INFN), Sezione di Roma, Piazzale Aldo Moro 2, 00185 Roma, Italy}

\affil{$^5$Southwestern Institute of Physics, P.O. Box 432, Chengdu 610041, People's Republic of China}

\affil{$^6$Key Laboratory of Frontier Physics in Controlled Nuclear Fusion and Institute of Plasma Physics, Chinese Academy of Sciences,
Hefei 230031, People’s Republic of China}

\affil{$^*$Author to whom any correspondence should be addressed.}

\email{fulvio.zonca@enea.it}


\begin{abstract}
We develop a general linear gyrokinetic theory of the low-frequency shear Alfvén wave and slow sound wave continuous spectra in toroidal plasmas, valid for arbitrary axisymmetric magnetic equilibria and equilibrium particle distribution functions within the adopted orderings. The kinetic continuous spectrum is interpreted as an intermediate asymptotic limit of the local plasma response, characterized by radial scales short compared with the poloidal wavelength but long compared with particle gyroradii and magnetic drift orbit widths. In ballooning space, quasineutrality and vorticity equations, closed by the drift/banana-center gyrokinetic response, form a coupled non-self-adjoint system of integro-differential equations with periodic coefficients and nonlocal kinetic kernels. An extended Bloch–Floquet formulation yields the renormalized plasma response in quadrature form, accounting for Landau damping, kinetic compressibility, finite parallel electric fields and toroidal multiscale structure. Retaining finite Larmor radius and magnetic drift orbit width effects extends the kinetic-continuum description to dispersive wave propagation and collisionless absorption, with the polarization of the different branches determined self-consistently. The theory illuminates the peculiar properties of low-frequency kinetic Alfvén waves in toroidal geometry and recovers their well-known high-frequency limit. Within the general fishbone-like dispersion-relation framework, the preferential excitation of Alfvénic polarization by energetic particles is explained in terms of both weaker collisionless damping and greater susceptibility to resonant energetic particle drive. The formulation provides a theoretical foundation for investigating the spectral properties underlying cross-scale interactions in toroidal plasmas.\end{abstract}

\section{Introduction}
\vskip 1em

At low frequencies comparable to characteristic thermal-particle frequencies, kinetic effects modify the 
compressional response in high-temperature toroidal plasmas, with important consequences for the stability of drift Alfv\'en waves (DAWs) 
driven by both thermal free energy and resonant energetic particles (EPs) \cite{chen16}. 
Using linear gyrokinetic theory, we analyze the shear Alfv\'en wave (SAW) and slow sound wave (SSW) continuous spectra and show how 
mode polarization and dispersion are altered by Landau damping and kinetic compressibility, including finite parallel electric field effects. 

We first formulate the low-frequency kinetic continuous-spectrum problem in ballooning space \cite{zonca14a,zonca14b,chen17,falessi19b}. 
This representation is particularly convenient because it captures the local structure of the fluctuations and their intrinsic multiscale character, 
including parallel mode structure and wave–particle resonances \cite{wei24,wei25,wei26}. The fundamental field variables are the shear Alfv\'en stream function
$$\delta\psi=-\frac{1}{c}\nabla_\parallel^{-1}\partial_t\delta A_\parallel,$$
which determines the perpendicular magnetic perturbation through the parallel vector potential, and the scalar potential $\delta\phi$. 
We work instead with the generalized parallel potential $\delta\phi_\parallel\equiv\delta\phi-\delta\psi$, for which
$$\delta E_\parallel=-\nabla_\parallel\delta\phi_\parallel.$$
Thus, the ideal-MHD condition $\delta E_\parallel=0$ corresponds, up to a field-line constant, to $\delta\phi_\parallel=0$. 
The quasineutrality condition determines the generalized parallel potential, while the vorticity equation determines the Alfv\'enic stream function 
and the associated generalized inertia and Floquet exponent. Both equations are closed by the gyrokinetic particle-compression response obtained from the drift/banana-center gyrokinetic equation, 
yielding a coupled integro-differential problem.

Floquet solutions \cite{falessi19b} can be constructed for general tokamak geometries, and the renormalized plasma response is expressed analytically 
in quadrature form. Floquet analysis of integro-differential equations with periodic coefficients and Volterra-type kernels has been developed in the literature \cite{belbas13}.
The present analysis extends these approaches 
to magnetized toroidal plasmas, where the relevant operators are generally non-self-adjoint and the solutions cannot, in general, 
be constructed from a finite-dimensional monodromy matrix because of the intrinsic spatial and temporal nonlocality of the kinetic response.

At short wavelengths the formulation recovers the kinetic Alfv\'en wave (KAW) dispersion and damping, providing a robust 
test case for gyrokinetic codes. This is particularly important because the generation of zonal electromagnetic 
fields depends sensitively on the spectral properties of the underlying fluctuations \cite{chen24,chen25}, including their KAW absorption and propagation characteristics \cite{zonca25}.

Within the general fishbone-like dispersion relation (GFLDR) framework \cite{zonca14a,zonca14b}, 
the theory provides a compact description of resonant low-frequency DAW excitation by EPs and explains the preferential excitation of Alfv\'enic polarization.

The plan of this work is as follows. Section 2 presents the theoretical formulation of the problem of interest. The concept of continuous spectrum and its physical meaning within the 
framework of gyrokinetic theory is presented in Section 3, explaining that it should be rigorously intended as asymptotic limit of the local plasma response. Consistently with its definition,
the Bloch-Floquet reduction of the kinetic problem remains valid, although the standard approach based on monodromy-matrix methods cannot be applied. The importance of mode polarization 
as characterization of the fluctuation structures is also introduced here, arguing its implications for the stability properties.
The properties of KAW in tokamak plasmas are discussed in Section 4, where it is shown that the general description recovers the well-known high frequency limit. There, 
the KAW propagation and absorption in the long wavelength limit are also presented. Section 5 provides final discussions and conclusions, while technical details regarding the 
derivation of KAW in the long wavelength limit and the specialization to well circulating particle responses in circular plasmas are given in two appendixes for interested readers.

\section{Theoretical formulation}
\vskip 1em

In this work we focus our attention on the low frequency continuous spectra, where $\omega$ is comparable to characteristic thermal-particle transit, bounce, and toroidal precession frequencies.
In fact, the parallel electric field in toroidal plasmas is due to either finite Larmor radius (FLR) of thermal ions or to kinetic compressional particle response. The latter becomes negligible for increasing 
frequency, since it scales as $\omega_d/\omega$, $\omega_d = \bm k \cdot \bm v_d$ with $\bm k$ the wave vector and $\bm v_d$ the magnetic drift velocity. Thus, peculiar behaviors connected with 
KAW propagation and absorption in toroidal geometry are expected to become increasingly more important at low frequency. Furthermore, this is also the frequency range where 
resonant low-frequency DAW excitation by EPs plays crucial roles in cross scale couplings between drift wave turbulence and Alfv\'enic fluctuations \cite{chen16,zonca15b}, with possible beneficial
impact of EPs on turbulent transport \cite{citrin23,garcia24,na25,mazzi26}.

We adopt straight magnetic field line flux coordinates $(r,\theta,\zeta)$, with $r$ a radial-like flux variable, $\theta$ the poloidal angle and $\zeta$ the generalized toroidal angle, which the equilibrium is independent of. 
Thus, the safety factor $q(\psi) = \bm B_0 \cdot \bm \nabla \zeta/\bm B_0 \cdot \bm \nabla \theta$ is a magnetic flux function, the
generic tokamak equilibrium magnetic field is given by $\bm B_0 = \bm \nabla (\zeta - q \theta) \times \bm \nabla \psi$ and the Jacobian ${\cal J}= (\bm \nabla \psi \times \nabla \theta \cdot \bm \nabla \zeta)^{-1}$ is such that
${\cal J} B_0^2$ is a flux function itself (Boozer coordinates). For $|\omega|^2\ll\omega_A^2$,  with $\omega_A = v_A/(q R_0)$ the Alfv\'en frequency, $v_A$ the Alfv\'en speed and $R_0$ the tokamak major radius, low frequency fluctuations 
in the ballooning space, denoted by the extended poloidal angle $\vartheta$, are characterized by two characteristic scales; a $\vartheta_0 \sim {\cal O}(1)$ scale reflecting equilibrium modulations in the poloidal angle,
and a long scale of variation, $\vartheta_1 \sim {\cal O}(\omega_A/|\omega|) \gg 1$, which describes the flute-like nature of underlying parallel mode structures; \idest, $|k_\parallel | q R_0 \ll 1$ \cite{zonca14a,zonca14b}.

Introduce now the ballooning space representation of $\delta \phi_\parallel$, 
$$\delta \phi_\parallel \mapsto |\nabla r| \delta \bar \Phi_{\parallel 0}/\hat \kappa_\perp , $$
with $\hat \kappa_\perp^2$ the ballooning space representation of $-\nabla_\perp^2/k_\vartheta^2$ and $k_\vartheta = -nq/r$, with $n$ the toroidal mode number. Similarly, let 
$$\delta \psi \mapsto |\nabla r| \delta \bar \Psi_{0}/\hat \kappa_\perp $$
be the ballooning space representation of $\delta \psi$. We also represent $\delta \bar \Phi_{\parallel 0}$ and $\delta \bar \Psi_{0}$ as asymptotic expansions in the perturbation parameter $|\omega|/\omega_A\ll 1$. Thus, we formally write
$$\delta \bar \Phi_{\parallel 0} = \delta \bar \Phi_{\parallel 0}^{(0)} + \delta \bar \Phi_{\parallel 0}^{(1)} + ... , $$
and 
$$\delta \bar \Psi_{0} = \delta \bar \Psi_{0}^{(0)} + \delta \bar \Psi_{0}^{(1)} + ... .$$
Following \cite{chen16,zonca14a,zonca14b}, it is possible to show that $\partial_{\vartheta_0}^2  \delta \bar \Psi_{0}^{(0)} = \partial_{\vartheta_0}^2  \delta \bar \Psi_{0}^{(1)} = 0$ up to the first two orders; \idest, that 
$\delta \bar \Psi_{0}^{(0)} (\vartheta_1)$ is a function of $\vartheta_1$ only, while $\delta \bar \Psi_{0}^{(1)} (\vartheta_1) = 0$ without loss of generality \cite{zonca96b}.
It is possible to show that, for $|\vartheta_1| \gg 1$, $\delta \bar \Psi_{0}^{(0)}$ satisfies the vorticity equation
\begin{equation}\partial_{\vartheta_1}^2 \delta \bar \Psi_{0}^{(0)} + \Lambda^2 \delta \bar \Psi_{0}^{(0)} = 0 \; , \label{eq:gkvort} \end{equation}
where, in the limit of vanishing FLR, $\lambda_L^2 = \hat \kappa_\perp^2 k_\vartheta^2 \mu B_0/\Omega^2 \rightarrow 0$, 
\begin{eqnarray}
& & \hspace*{-3em} \Lambda^2 = \lim_{\lambda_L \rightarrow 0} \overline{\left[\frac{\omega^2 {\cal J}^{2} B_{0}^{2}}{v_A^2} \left(1 - \frac{\omega_{*pi}}{\omega} \right) \left( 1 + \delta \bar \Phi_{\parallel 0}/\delta \bar \Psi_0^{(0)}\right)\right]_{\vartheta_0}}  + \sum \int {\cal E} d {\cal E} d \lambda  \label{eq:Lambda2} \\
& & \times \sum_\sigma \tau_b \overline{\left[\frac{4\pi  \omega {\cal J} B_0/\bar B_0}{k_\vartheta c} \kappa_g \sgn (s \vartheta_1)  \frac{
m (\mu B_0 + v_\parallel^2 ) J_0 (\lambda_L) \delta \hat K}{|\bm \nabla r| \delta \bar \Psi_{0}^{(0)}} \right]}\; . \nonumber 
\end{eqnarray}
is the generalized inertia. Here, for simplicity, we have considered one single ion species\footnote{The case of multiple ion species can be recovered with some additional formal complication and without qualitatively altering the present approach.},
various symbols are introduced as the magnetic moment $\mu = v_\perp^2/(2B_0)$, the cyclotron frequency $\Omega = e B_0/(m c)$ for particles of charge $e$ and mass $m$, 
the thermal ion diamagnetic frequency $\omega_{*pi}$, $\kappa_g = (R B_\phi/|\nabla \psi|) \partial_\theta B_0/({\cal J} B_0^2)$ is the geodesic curvature, $R B_\phi$ is a magnetic flux function with $B_\phi$ the equilibrium toroidal magnetic field, 
$s = r d \ln q/dr$ is the magnetic shear,  $\bar B_0$ is the equilibrium magnetic field value on axis, and we have introduced velocity space variables ${\cal E} = v^2/2$ and $\lambda = \mu \bar B_0/{\cal E}$,
with $J_0(\lambda_L)$ denoting the Bessel function accounting for FLR effects.
Furthermore, $\sum$ with no index/subscripts denotes summation on particle species, $\tau_b = \oint d \theta/\dot \theta$ is the particle bounce/transit time on a closed poloidal orbit, $\overline{(...)} = \tau_b^{-1} \oint d \theta(...)/\dot \theta$ denotes orbit averaging along particle equilibrium orbits, summation on $\sigma = v_\parallel/|v_\parallel|$ applies to circulating particles only,  and $\overline{\left( ... \right)_{\vartheta_0}}$ denotes $\vartheta_0$-scale averaging.
As a result of Eq. (\ref{eq:gkvort}), $\delta \bar \Psi_{0}^{(0)}\sim e^{i \Lambda |\vartheta_1|}$ asymptotically \cite{zonca14a,zonca14b}. Note that the dependence on $|\vartheta_1|$ in the exponential is consistent with causality constraints and yields 
to the general fishbone-like dispersion relation (GFLDR) when proper matching of the large-$|\vartheta_1|$ solution is imposed with the $\vartheta_0$-scale averaged potential, which vanishes for large $|\vartheta_1|$; \idest, 
\begin{equation}
i \Lambda = \delta \bar W_f + \delta \bar W_k \; , \label{eq:GFLDR}
\end{equation}
with $\delta \bar W_f$ and $\delta \bar W_k$ the fluid and kinetic contribution to the normalized potential energy, respectively \cite{zonca14a,zonca14b}. 
Equations (\ref{eq:gkvort}) and (\ref{eq:Lambda2}) illuminate the physical meaning of kinetic continuous spectrum being characterized, on the one hand, by sufficiently
long scales with respect to the thermal ion FLR ($\lambda_L \rightarrow 0$); and, on the other hand, by sufficiently short scales with respect to the poloidal wavelength of the considered fluctuation ($|s\vartheta_1| \sim \hat \kappa_\perp/|\bm \nabla r|
\rightarrow \infty$) \cite{zonca14a,zonca14b}.
This is clearly a conceptual abstraction that we will overcome when addressing KAW in Section 4. It is also a clear definition of the concept of kinetic continuous spectrum as asymptotic limit of the local plasma response, connected with 
physical spatial scales. Introduction of KAW in Section 4 will also illuminate the temporal scales involved with the kinetic continuous spectrum.

Equation (\ref{eq:gkvort}) is closed by the kinetic particle compressional response, $\delta \hat K$, which is most conveniently expressed in terms of the magnetic-drift/banana-center response $\delta \hat K_B = e^{i \hat Q_B} \delta \hat K$ \cite{wei25,wei26,tsai93,chen94}, with 
$\hat Q_B \simeq R B_\phi/(d\psi/dr)$ $(k_\vartheta v_\parallel/\Omega) s \vartheta_1$ asymptotically, for $|\bm \nabla r|/\hat \kappa_\perp \sim | s \vartheta_1|^{-1}$ $\sim {\cal O}(|\omega|/\omega_A) \ll 1$,
describing finite magnetic drift orbit width effects (FOW). Meanwhile, $\delta \hat K_B$ is obtained from the solution of the  magnetic-drift/banana-center gyrokinetic equation
\begin{equation}
\left( \omega_b \partial_{\theta_c} - i \omega + i n \bar \omega_d \right) \delta \hat K_B = 
 i \frac{e}{m}  e^{i \hat Q_B} J_0(\lambda_L) Q \bar F_0  \left[ \frac{|\nabla r|}{\hat \kappa_\perp} \delta \bar \Phi_{\parallel 0} + \frac{\omega_d}{\omega}  \frac{|\nabla r|}{\hat \kappa_\perp}   \delta \bar \Psi_{0}^{(0)} \right] \; , \label{eq:dKgke}
\end{equation}
where the canonical angle $\theta_c$ maps into the ballooning angle $\vartheta$ via the definition $\dot \vartheta \partial_\vartheta = \omega_b \partial_{\theta_c}$\cite{zonca15a}, $\omega_b = 2\pi/\tau_b$ is the transit/bounce frequency, $\bar 
\omega_d$ is the toroidal precession frequency, $\omega_d \simeq (\mu B_0 + v_\parallel^2) (-k_\vartheta/\Omega) \kappa_g \hat \kappa_\perp \sgn (s\vartheta_1)$ is the asymptotic expression of the magnetic drift frequency 
for large $|s \vartheta_1|$, and 
\begin{equation}
Q \bar F_0 = - \frac{m}{T_0} \omega \left[ 1 - \frac{\omega_{*n}}{\omega} - \frac{\omega_{*T}}{\omega} \left( \frac{m {\cal E}}{T_0} - \frac{3}{2} \right) \right] \bar F_0
\; \label{eq:QbarF0Max}
\end{equation}
for the case of a (canonical) Maxwellian distribution function that will be of interest here, with $T_0$ the local equilibrium temperature, and 
$\omega_{*n}$ and $\omega_{*T}$ the diamagnetic frequencies due to density and temperature gradients, respectively; \idest,
$\omega_{*p} = \omega_{*n}+\omega_{*T}$. The general solution of Eq. (\ref{eq:dKgke}) can be written as
\begin{eqnarray}
\delta \hat K_B & = & \left\{\sum_{p \in \mathbb Z} \frac{e}{m} \overline{\frac{e^{i \hat Q_B} Q\bar F_0 (J_0 |\bm \nabla r| \delta \bar \Phi_{\parallel 0}/\hat \kappa_\perp) e^{-i \Lambda {\rm sgn}(\vartheta_1) \bar \theta_c 
 -ip\theta_c}}{n\bar\omega_{d} + p \omega_b + \Lambda \dot{\bar\theta}_c {\rm sgn}(\vartheta_1) - \omega}  }
 \right. \label{eq:fdklow}\\
 & & \hspace{-6em} \left. + \sum_{p \in \mathbb Z} \frac{e}{m} \overline{\frac{e^{i \hat Q_B} Q\bar F_0 (\omega_d/\omega) (J_0|\bm \nabla r| \delta \bar \Psi_{0}^{(0)}/\hat \kappa_\perp) e^{-i \Lambda {\rm sgn}(\vartheta_1) \bar \theta_c
 -ip\theta_c}}{n\bar\omega_{d} + p \omega_b + \Lambda \dot{\bar\theta}_c {\rm sgn}(\vartheta_1) - \omega}  } \right\}
 e^{i \Lambda {\rm sgn}(\vartheta_1) \bar \theta_c  + i p \theta_c}  
 \;\; , \nonumber 
\end{eqnarray} 
where the extended ballooning angle has been mapped into $\theta_c$ as $\vartheta \equiv \bar \theta_c + \tilde \Theta(\theta_c)$, $\dot{\bar\theta}_c = \sigma \omega_b$ for circulating particles and vanishes otherwise, 
and $\tilde \Theta(\theta_c)$ is a periodic function of $\theta_c$ vanishing at $\theta_c = 2\pi \ell$ with $\ell \in \mathbb Z$ \cite{wei25,wei26,zonca15a}. Equation (\ref{eq:fdklow}) is the general solution of Eq. (\ref{eq:dKgke})
given as expansion in bounce/transit harmonics of $\theta_c$.

Following \cite{chen16,zonca14a,zonca14b}, Eqs. (\ref{eq:gkvort}) and (\ref{eq:Lambda2}) are closed by the quasineutrality condition
\begin{equation}
\sum \frac{n_{0} e^2}{T_{0}} \frac{|\nabla r|}{\hat \kappa_\perp} \delta \bar \Phi_{\parallel 0} =  \sum_{\ell \in \mathbb Z} e^{i \ell \vartheta_0} \sum \int {\cal E} d {\cal E} d \lambda \tau_b  \sum_\sigma
\overline{\left[ e^{- i \ell \vartheta_0} \frac{e J_0(\lambda_L)\delta \hat K}{{\cal J} \bar B_0}\right]}  \;\; . \label{eq:quasineutrality}  
\end{equation}
These equations form a closed system that allows computing the low frequency kinetic continuous spectrum in general toroidal geometry, when Eq. (\ref{eq:fdklow}) is substituted back into them 
in the limit $\lambda_L^2, \hat Q_B^2 \rightarrow 0$, $\hat \kappa_\perp^2 \rightarrow \infty$; \idest, neglecting FLR and FOW effects. This will be investigated in more detail in the next Section. Meanwhile the role of FOW and FLR in KAW excitation in toroidal geometry will be addressed in Section 4.

\section{Kinetic continuous spectrum, Floquet solutions and polarization}
\vskip 1em 

Equations (\ref{eq:gkvort}), (\ref{eq:Lambda2}), and
(\ref{eq:quasineutrality}), together with the gyrokinetic response in
Eq.~(\ref{eq:fdklow}), define a closed integro-differential problem for
the kinetic continuous spectrum. In the limit
$\lambda_L^2,\hat Q_B^2\rightarrow 0$ and
$\hat\kappa_\perp^2\rightarrow\infty$, the coefficients and
kinetic-response kernels become $2\pi$-periodic in the ballooning
coordinate within each asymptotic region of large positive or negative
$\vartheta_1$.\footnote{The global matching between the large-$|\vartheta_1|$
asymptotic regions is imposed by the GFLDR, Eq.~(\ref{eq:GFLDR}), as
discussed in Sec.~2.} The corresponding
kinetic-response kernels inherit the same periodic-lattice structure,
in the sense of invariance under simultaneous translation of their
ballooning-space arguments by $2\pi$. As anticipated in Section 2, the kinetic continuous spectrum
is intended as asymptotic limit of the local plasma response, implied by the condition $\hat\kappa_\perp^2\rightarrow\infty$, 
subject to the constraint of neglecting FLR and FOW effects as expressed by $\lambda_L^2,\hat Q_B^2\rightarrow 0$.
It is worthwhile noting that spatial locality is embedded in the intrinsic one-dimensional nature of the problem of interest, which,
from tokamak geometry, can be extended to general stellarator equilibria under the condition of near integrability of the 
equilibrium magnetic field lines. In that case, it is possible to show that the periodic-lattice structures, typical of tokamaks,
is replaced by quasiperiodic-lattice strucures of stellarators. The novel features that are implied by this property will be
discussed in a separate work and will not be mentioned any further in the present study.

For definiteness, consider the asymptotic region
$\vartheta_1\rightarrow+\infty$. The solutions can then be decomposed
into Bloch sectors. In particular, a Bloch-sector solution satisfies the quasiperiodicity condition
\begin{equation}
\begin{pmatrix}
\delta\Psi_0(\vartheta_0+2\pi)\\
\delta\Phi_{\parallel 0}(\vartheta_0+2\pi)
\end{pmatrix}
=
e^{2\pi i\Lambda}
\begin{pmatrix}
\delta\Psi_0(\vartheta_0)\\
\delta\Phi_{\parallel 0}(\vartheta_0)
\end{pmatrix},
\label{eq:floquettrans}
\end{equation}
where $e^{2\pi i\Lambda}$ is the Floquet multiplier and $\Lambda$ is
the corresponding characteristic exponent. An analogous construction
applies in the asymptotic region
$\vartheta_1\rightarrow-\infty$, with the global connection between the
two regions determined by the GFLDR.

The Bloch--Floquet reduction remains applicable because the limiting
problem retains its periodic-lattice structure, despite the
nonlocality of the kinetic response. The analysis can therefore be
restricted to a single Bloch sector of the $2\pi$-periodic ballooning
lattice, with $\Lambda$ serving as the associated characteristic
exponent.

This kinetic Floquet problem differs fundamentally from the
corresponding ideal-MHD problem. In the ideal-MHD limit, the continuous
SAW and SSW spectra can be formulated in terms of ordinary differential
equations and analyzed using standard monodromy-matrix methods
\cite{falessi19b,wei24}. In the kinetic problem,
resonant particle dynamics and Landau damping make the operator
non-self-adjoint, and the characteristic exponent can consequently
become complex. Moreover, because the kinetic response is represented
by nonlocal integro-differential kernels, the solution is not generally
determined by a finite-dimensional set of local initial data.
Consequently, the kinetic Floquet problem cannot, in general, be
reduced to a finite-dimensional monodromy matrix, as is possible for
the Hill equation describing the simplest SAW continuum or for the
corresponding coupled SAW--SSW system
\cite{falessi19b,wei24}.

All subsequent expressions in this section are evaluated in the limit $\lambda_L^2,\hat Q_B^2\rightarrow 0$.
Substituting Eq.~(\ref{eq:fdklow}) into the quasineutrality condition Eq. (\ref{eq:quasineutrality}), we obtain
\begin{eqnarray}
& & A_{l,l'}  \left( |\bm \nabla r| \delta \bar \Phi_{\parallel 0}/\hat \kappa_\perp \right)_{l'} = C_l \left( |\bm \nabla r| \delta \bar \Psi_{0}^{(0)}/\hat \kappa_\perp \right) \; , \nonumber \\
& & A_{l,l'}  = - \sum \int {\cal E} d {\cal E} d \lambda \sum_\sigma \tau_b  \sum_{p \in \mathbb Z} \frac{T_{0i}}{n_0 m}    Q\bar F_{0} \overline{  e^{i l'\vartheta_0} e^{ -ip\theta_c} } \nonumber \\ & & \hspace*{3em} \times \overline{\frac{\left( {\cal J} \bar B_0 \right)^{-1} 
e^{ i p \theta_c}   e^{-il\vartheta_0} }{n\bar\omega_{d} + p \omega_b + \Lambda \dot{\bar\theta}_c {\rm sgn}(\vartheta_1) - \omega} }
+  \left( 1 + \frac{1}{\tau} \right) \delta_{l,l'}  , \label{eq:qnsolvelow} \\
& & C_l = \sum \int {\cal E} d {\cal E} d \lambda \sum_\sigma \tau_b  \sum_{p \in \mathbb Z} \frac{T_{0i}}{n_0 m}    Q\bar F_{0}   \overline{ (\omega_d/\omega) e^{ -ip\theta_c}} \nonumber \\ & & \hspace*{3em} \times \overline{\frac{\left( {\cal J} \bar B_0 \right)^{-1} 
e^{ i p \theta_c}   e^{-il\vartheta_0} }{n\bar\omega_{d} + p \omega_b + \Lambda \dot{\bar\theta}_c {\rm sgn}(\vartheta_1) - \omega}  } \; . \nonumber
\end{eqnarray}
Here, repeated Fourier indices are summed over $\mathbb Z$ unless otherwise
indicated, $\left( |\bm \nabla r| \delta \bar \Phi_{\parallel 0}/\hat \kappa_\perp \right)_l$ denote the $l$-Fourier harmonic in $\vartheta_0$ of the considered field and $\tau = T_{0e}/T_{0i}$. 
It is worthwhile noting that 
\begin{equation}
{\bm \epsilon}_s \equiv \tau A_{l,l'} \; \label{eq:epsilons}
\end{equation}
is the projection of the electrostatic drift wave dispersion tensor on the Fourier basis in the vanishing FLR limit. 
Specializing this result to $|\omega| \ll \omega_{bi}$, we can estimate \cite{chavdarovski09} 
\begin{equation}
\left( |\bm \nabla r| \delta \bar \Phi_{\parallel 0}/\hat \kappa_\perp \right)_l
\sim f_t  (\omega_{di} \omega)/\omega_{bi}^2 \left( |\bm \nabla r| \delta \bar \Psi_{0}^{(0)}/\hat \kappa_\perp \right) \; ,  \label{eq:eparestimate}
\end{equation}
with $f_t$ the trapped particle fraction; which implies that the nonconstant Fourier harmonics of the parallel
electric-field component are negligible at this order, consistently
with the ideal-MHD limit. Meanwhile, $\left( |\bm \nabla r| \delta \bar \Phi_{\parallel 0}/\hat \kappa_\perp \right)_0$ is finite and given by
\begin{eqnarray}
& & A_{0,0}  \left( |\bm \nabla r| \delta \bar \Phi_{\parallel 0}/\hat \kappa_\perp \right)_{0} = C_0 \left( |\bm \nabla r| \delta \bar \Psi_{0}^{(0)}/\hat \kappa_\perp \right) \; , \nonumber \\
& & A_{0,0}  =  \left( 1 + \frac{1}{\tau} \right)+ \sum \int {\cal E} d {\cal E} d \lambda \sum_\sigma \tau_b \frac{T_{0i}}{n_0 m}  \overline{ \left( {\cal J} \bar B_0 \right)^{-1}} \frac{ Q\bar F_{0}  (\omega - n \bar\omega_d) }{(\omega - n\bar\omega_{d})^2 -  \Lambda^2 \dot{\bar\theta}_c^2} \; ,  \label{eq:qnsolvel0} \\
& & C_0 = - \sum \int {\cal E} d {\cal E} d \lambda \sum_\sigma \tau_b \frac{T_{0i}}{n_0 m}  \overline{ \left( {\cal J} \bar B_0 \right)^{-1}} \frac{ (Q\bar F_{0}/\omega) n \bar\omega_d   (\omega - n \bar\omega_d) }{(\omega - n\bar\omega_{d})^2 -  \Lambda^2 \dot{\bar\theta}_c^2} \; . \nonumber
\end{eqnarray}
This means that both magnetically trapped electrons and ions can contribute to the violation of the ideal MHD condition $\left( |\bm \nabla r| \delta \bar \Phi_{\parallel 0}/\hat \kappa_\perp \right)_{0} = 0$. 
This is connected with the non cancellation of precessional drifts at low frequency. However, when mode frequency becomes significantly larger than the precession frequency, the trapped particle contribution becomes negligible. 
Note that EPs can also contribute to this process if their energy is not too high, so that their density response in the quasineutrality condition is not ignorable. 

Strictly speaking, the estimate in Eq.~(\ref{eq:eparestimate}) applies
to $v_\parallel$-symmetric particle equilibrium distribution functions and up-down symmetric magnetic equilibria, for which, for example,
$B_0$ is an even function of $\vartheta_0$. In a generic equilibrium,
breaking $v_\parallel$ particle distribution symmetry and/or up-down magnetic equilibrium symmetry can produce a finite parallel electric
field through geodesic-curvature effects. This effect scales as
$\sim k_\vartheta s\vartheta_1\delta\bar\Psi_0^{(0)}$ and results from
the incomplete cancellation of radial magnetic-drift contributions\footnote{This is consistent with the integrability of equilibrium particle motion
and with the existence of closed particle orbits in the poloidal cross
section of an axisymmetric tokamak.}. 
The corresponding $l$-Fourier harmonics of
$\left(|\bm\nabla r|\delta\bar\Phi_{\parallel 0}/\hat\kappa_\perp\right)_l$
scale as $\sim\omega_{di}/\omega_{bi}$, as follows from
Eq.~(\ref{eq:qnsolvelow}), rather than as
$\sim\omega_{di}\omega/\omega_{bi}^2$ as in Eq.~(\ref{eq:eparestimate}).
This contribution remains small and is associated with finite-orbit-width
effects.

\begin{figure}
 \centering
        \includegraphics[width=0.8\textwidth]{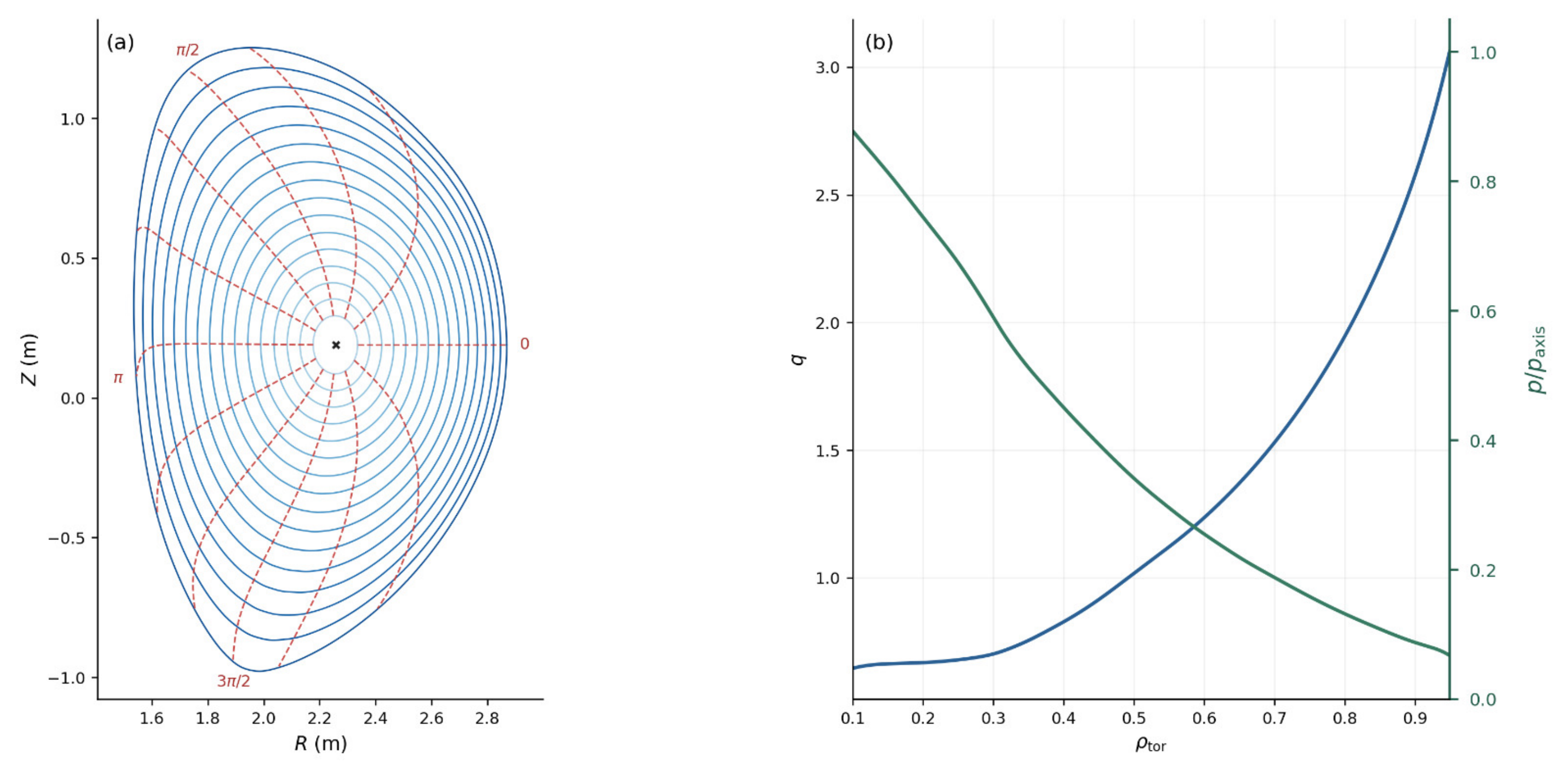}
 \caption{Reference DTT magnetic equilibrium \cite{crisanti24,giruzzi26} adopted in Fig. \ref{fig2}  for illustrating the behavior of SAW and SSW continuous spectra vs. the parallel wave-vector. (a) Magnetic flux surfaces are shown along with Boozer coordinates.
 (b) Equilibrium pressure and $q$ profiles are shown vs. $\rho_{tor}$ defined 
 as the square root of the normalized toroidal flux.}
\label{fig1}
\end{figure}

Following the same procedure, substitution of Eq.~(\ref{eq:fdklow}) into the generalized inertia expression Eq. (\ref{eq:Lambda2}) yields
\begin{eqnarray}
& & \hspace*{-3em} \Lambda^2 = \overline{\left[\frac{\omega^2 {\cal J}^{2} B_{0}^{2}}{v_A^2} \left(1 - \frac{\omega_{*pi}}{\omega} \right) \left( 1 + C_0/A_{0,0} \right) \right]_{\vartheta_0}}  \nonumber \\
& & - \frac{4\pi  \omega^2 {\cal J} B_0^2/\bar B_0}{k_\vartheta^2 c^2} \sum \int {\cal E} d {\cal E} d \lambda  \sum_\sigma \tau_b \sum_{p \in \mathbb Z}  Q\bar F_0 \frac{e^2}{m} \overline{\left[ \frac{\left(\omega_d/\omega\right) e^{i p \theta_c}}{\hat \kappa_\perp |\bm \nabla r|}   \right]}  \label{eq:Lambda2gen} \\
& &  \times \overline{\left[ \frac{\left( |\bm \nabla r| /\hat \kappa_\perp \right) e^{ - i p \theta_c}  \left( \omega_d/\omega + e^{i l \vartheta_0} (A^{-1})_{l,l'} C_{l'} \right) }{n\bar\omega_{d} + p \omega_b + \Lambda \dot{\bar\theta}_c {\rm sgn}(\vartheta_1) - \omega} \right]}\; . \nonumber 
\end{eqnarray}
Here, we have used the fact that we are working in Boozer coordinates to extract ${\cal J} B_0^2$ from the orbit integral. Furthermore, we have considered only the parallel electric field contribution from Eq. (\ref{eq:qnsolvel0}), since it is the only contribution that can be ${\cal O}(1)$ in 
Eq. (\ref{eq:Lambda2gen}). Meanwhile, formally, $C_l \sim \hat \kappa_\perp$ for $l \neq 0$.
Thus, Eq. (\ref{eq:Lambda2gen}) is independent of $\hat \kappa_\perp$ in the limit $\hat \kappa_\perp\rightarrow \infty$, and provides the expression for the generalized inertial at low-frequency, $|\omega|\ll \omega_A$, which applies for 
general tokamak plasma equilibria, including non up-down symmetric ones. Equation (\ref{eq:Lambda2gen}) can be used as a definition of the kinetic continuous spectrum given by \cite{zonca14a,zonca14b}
\begin{equation}
\Lambda^2 = \overline{\left( k_\parallel^2 {\cal J}^2 B_0^2 \right)_{\vartheta_0}} \; . \label{eq:kincon}
\end{equation}

\begin{figure}
 \centering
        \includegraphics[width=0.8\textwidth]{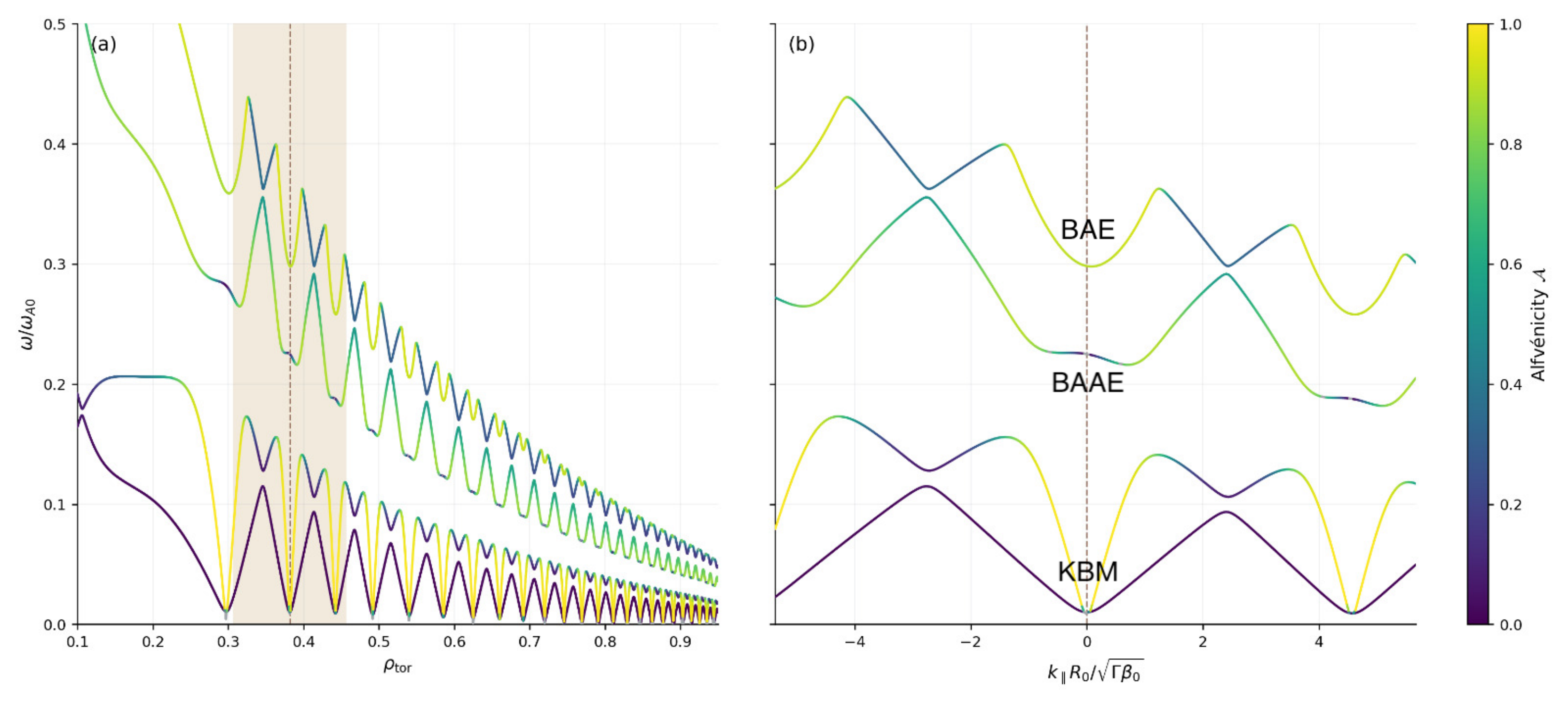}
 \caption{(a) Low-frequency SAW and SSW continuous spectra for $n=10$ for the reference DTT magnetic equilibrium of Fig. \ref{fig1}. The continuous spectra are represented for $\omega/\omega_{A0}$ in the MHD limit
 solely for illustration, with $\omega_{A0} = v_{A0}/R_0$ and  $v_{A0}$ the Alfv\'en velocity on axis. A similar figure can be obtained for the kinetic case, as shown in Ref. \cite{li23} (cf. Fig. 6 therein) and in Appendix B (cf. Fig. \ref{fig3}).
 (b) The KBM, BAAE and BAE branches are clearly identified. The shaded area in the left figure is expanded and plotted as a function of $k_\parallel R_0/(\Gamma^{1/2} \beta^{1/2})$, with $\Gamma$ the ratio
 of specific heats. Lines are colored and reflect the value of Alfv\'enicity, defined in Eq. (\ref{eq:alfvenicity}), as reported in the color palette.}
 \label{fig2}
\end{figure}

These expressions are generally valid and
are applicable to long wavelength drift Alfv\'en waves (DAW). Equations (\ref{eq:qnsolvelow}) and (\ref{eq:Lambda2gen} )
show that one can solve for the low frequency fluctuation spectrum by means of quadratures involving equilibrium quantities only. These results generalize earlier ones \cite{zonca96b,chavdarovski09,zonca10,chavdarovski14} to arbitrary tokamak equilibria and particle distribution functions.
In particular, they recover the expressions of generalized inertia in the limit of well-circulating thermal ions \cite{zonca96b,zonca10} and well-circulating and deeply trapped particles \cite{chavdarovski09,chavdarovski14}. 
A key result remains that the strength of Landau damping is connected with the role of $\delta \bar \Phi_{\parallel 0}$; \idest, with the mode polarization, which can be generally defined as a scalar quantity that measures the relative strength of the components in the solution of Eq. (\ref{eq:floquettrans}).
Possible definitions of polarization are the Alfv\'enicity \cite{falessi19b}
\begin{equation}
{\cal A} = \frac{\overline{\left( |\delta \bar \Psi_{0}^{(0)}|^2 \right)_{\vartheta_0}}}{\overline{\left( |\delta \bar \Psi_{0}^{(0)}|^2 \right)_{\vartheta_0}}+\overline{\left( |\delta \bar \Phi_{\parallel 0}|^2 \right)_{\vartheta_0}}} \; ; \label{eq:alfvenicity}
\end{equation}
or the acoustic polarization \cite{heidbrink25}
\begin{equation}
P_A  = \frac{\overline{\left( |\delta \bar \Phi_{\parallel 0}|^2 \right)_{\vartheta_0}}^{1/2}}{\overline{\left( |\delta \bar \Psi_{0}^{(0)}|^2 \right)_{\vartheta_0}}^{1/2}+\overline{\left( |\delta \bar \Phi_{\parallel 0}|^2 \right)_{\vartheta_0}}^{1/2}} \; . \label{eq:acpol}
\end{equation}
The values of ${\cal A}$ and $P_A$ are $\in [0,1]$ by definition, and need to be determined self-consistently after solving GFLDR and corresponding mode structures. 
Higher $P_A$, or equivalently lower ${\cal A}$, corresponds to stronger
Landau damping \cite{chavdarovski09,zonca10,chavdarovski14}. This
theoretical prediction has been supported by comparisons with
experimental observations \cite{ma22,xu22,ma23} and helps explain why
EP-driven resonant excitation preferentially favors Alfv\'enic
polarization. Another reason is connected with the structure of the GFLDR. From a comparison of Eqs. (\ref{eq:GFLDR}) and (\ref{eq:kincon}),
one readily recognizes that $\delta \bar W_f + \delta \bar W_k$ can be considered as an ``effective parallel wave-vector'' for physical fluctuations excited, \exgra, by EPs. Meanwhile, Fig. \ref{fig2} illustrates the behaviors
of kinetic ballooning mode (KBM) and beta-induced Alfv\'en Eigenmode (BAE) continuous spectra (predominantly Alfv\'enic) vs. $k_\parallel R_0/(\Gamma^{1/2} \beta^{1/2})$
compared with the corresponding behavior of the beta-induced Alfv\'en acoustic Eigenmode (BAAE) continuum (mixed polarization). To illustrate the changing polarization
of the different branches with frequency, the line color reflects the value of Alfv\'enicity as specified in the color palette. It is evident that the BAAE continuum has a much weaker dependence
on $k_\parallel R_0/(\Gamma^{1/2} \beta^{1/2})$ than the KBM/BAE continua. Correspondingly, the BAAE response to resonant excitation by EPs is weaker than that of 
fluctuations with Alfv\'enic polarization, as originally noted in \cite{chen17}.

\section{Kinetic Alfv\'en Waves in tokamak plasmas}
\vskip 1em

In order to explore the properties of KAW in tokamak plasmas, we need to take into account FLR and FOW effects in the vorticity equation, Eq. (\ref{eq:gkvort}), which can be written as \cite{zonca14a,chen01}
\begin{eqnarray} \partial_{\vartheta_1}^2 \delta \bar \Psi_{0}^{(0)} & +  &\overline{\left\{\frac{\omega^2 {\cal J}^{2} B_{0}^{2}}{v_A^2} \left[ \left(1 - \frac{\omega_{*ni}}{\omega} \right) 
\frac{1-\Gamma_0(b)}{b}  - \frac{\omega_{*Ti}}{\omega} \left( \Gamma_0(b) - \Gamma_1(b)\right) \right] \left( \delta \bar \Psi_0^{(0)} + \delta \bar \Phi_{\parallel 0}\right) \right\}_{\vartheta_0}} 
\nonumber \\ & - & \sum \int {\cal E} d {\cal E} d \lambda  \sum_\sigma \tau_b \overline{\left[\frac{4\pi  \omega {\cal J} B_0^2/\bar B_0}{k_\vartheta^2 c^2} e \frac{\omega_d}{\hat \kappa_\perp |\bm \nabla r|}   J_0 (\lambda_L) \delta \hat K \right]}
= 0 \; . \label{eq:gkvortKAW}
\end{eqnarray}
Here, $\Gamma_{0,1} (b) = e^{-b} I_{0,1}(b)$, $I_{0,1}(b)$ are modified Bessel functions of the first kind and $b = \hat \kappa_\perp^2 k_\vartheta^2 (T_{0i}/m_i)/\Omega_i^2$, while we still have $\delta \hat K = e^{-i \hat Q_B} \delta \hat K_B$, with 
$\delta \hat K_B$ given by Eq. (\ref{eq:fdklow}). Meanwhile, the quasineutrality condition becomes  \cite{zonca14a,chen01}
\begin{eqnarray}
 & &  \hskip - 2em \left( 1  + \frac{1}{\tau} \right) \frac{|\nabla r|}{\hat \kappa_\perp} \delta \bar \Phi_{\parallel 0}  +  \left[ \left(1 - \frac{\omega_{*ni}}{\omega} \right) \left(
1-\Gamma_0(b) \right) - \frac{\omega_{*Ti}}{\omega} b \left( \Gamma_0(b) - \Gamma_1(b)\right) \right]  \frac{|\nabla r|}{\hat \kappa_\perp}  \delta \bar \Psi_0^{(0)} = \nonumber \\
  & &  = \frac{T_{0i}}{n_0e^2} \sum_{\ell \in \mathbb Z} e^{i \ell \vartheta_0} \sum \int {\cal E} d {\cal E} d \lambda \sum_\sigma  \tau_b 
\overline{\left[ e^{- i \ell \vartheta_0} \frac{e J_0(\lambda_L)\delta \hat K}{{\cal J} \bar B_0}\right]}  \;\; , \label{eq:qnKAW}  
\end{eqnarray}
with $\tau = T_{0e}/T_{0i}$. Equations (\ref{eq:gkvortKAW}) and (\ref{eq:qnKAW}), supplemented by vanishing boundary conditions on $\delta \bar \Psi_0^{(0)}$ and $\delta \bar \Phi_{\parallel 0}$ as $\hat \kappa_\perp^2 \rightarrow \infty$,  
form a closed set of well posed equations for KAW in tokamak plasmas for arbitrarily short wavelength. As such, they 
allow us to explore the behavior of continuous spectra as $b \rightarrow \infty$ and generalize the problem of ``influence of gyroradius and dissipation on the Alfv\'en wave continuum'' originally posed by
Connor, Tang and Taylor for a magnetized plasma slab \cite{connor83}. In particular, it is possible to explore the transition from solutions of the discretized spectrum due to FLR/FOW effects to genuine continuous spectrum solutions
connected with the singularity at $\omega=0$. 

The general structure of these equations suggests that toroidal geometry introduces an important novelty in Eq. (\ref{eq:gkvortKAW}), which  
modifies both KAW propagation and absorption due to  geodesic curvature coupling. Without geodesic curvature, such as in a dipole field, the last term in Eq. (\ref{eq:gkvortKAW}) would be suppressed and
KAW properties would be determined by the parallel electric field, $\delta \bar \Phi_{\parallel 0}$, due to finite particle kinetic compression in Eq. (\ref{eq:qnKAW}). As a result, 
KAW behavior would still be modified by geometry via the precessional motion of trapped particles but, except for that, it would closely follow the well known description in a plasma slab \cite{hasegawa75,hasegawa76a,chen91,chen21}.
Another property of Eqs. (\ref{eq:gkvortKAW}) and (\ref{eq:qnKAW}), anticipated in Section 3 with the discussion of the concept of kinetic continuous spectrum,  
is that they are posed in the form of a one dimensional problem, notwithstanding their intrinsic integro-differential nature due to the
spatiotemporal nonlocality of the particle response, embedded in Eq. (\ref{eq:fdklow}). This makes them suitable of numerical solution with high accuracy in phase space for general tokamak equilibria, 
resolving for parallel mode structures and for wave-particle resonances responsible of collisionless dissipation. For this reason, Eqs. (\ref{eq:gkvortKAW}) and (\ref{eq:qnKAW}) may offer a useful 
test case for benchmarking gyrokinetic codes and their ability of reproducing KAW propagation and absorption.

\subsection{High frequency limit and recovery of KAW in a plasma slab}
\vskip 1em

An interesting limit to investigate to explore these properties is $|\omega_{de}| \sim |\omega_{di}| \sim \omega_{bi} \ll |\omega| \ll \omega_{be}$. For such a case, the FLR effects are fully retained but FOW effects are suppressed because 
of the frequency ordering with respect to the thermal ion bounce motion. Qualitatively, thermal ions do not have time to experience/recognize their magnetic drift motion since the response of thermal ions is 
given by 
\begin{equation}
\delta \hat K_i \simeq  - \left\{ \frac{e}{m} \frac{Q \bar F_0}{\omega} \left( \frac{|\nabla r|}{\hat \kappa_\perp} J_0(\lambda_L) \delta \bar \Phi_{\parallel 0}  \right) \right\}_i \; . \label{eq:dKfluidion}
\end{equation}
Meanwhile, thermal electrons are well described by their bounce averaged response
\begin{equation}
\delta \hat K_e \simeq  \left\{ \frac{e}{m} \frac{Q \bar F_0}{\omega} \overline{\left( \frac{|\nabla r|}{\hat \kappa_\perp} \delta \bar \Phi_{\parallel 0}  \right)_{\vartheta_0}} \frac{\omega}{\Lambda \dot{\bar\theta}_c {\rm sgn}(\vartheta_1)  - \omega} \right\}_e \; . \label{eq:dKeleave}
\end{equation}
When substituted back into the quasineutrality condition, Eq. (\ref{eq:qnKAW}), these expressions yield
\begin{eqnarray}
 & &  \hskip - 2em \left( \frac{1}{\tau} + \frac{\omega_{*ni}}{\omega} \right) \frac{|\nabla r|}{\hat \kappa_\perp} \delta \bar \Phi_{\parallel 0} +  \left[ \left(1 - \frac{\omega_{*ni}}{\omega} \right) \left(
1-\Gamma_0(b) \right) \right.  \label{eq:qnKAWlim} \\ & & \left. - \frac{\omega_{*Ti}}{\omega} b \left( \Gamma_0(b) - \Gamma_1(b)\right) \right]  \frac{|\nabla r|}{\hat \kappa_\perp} \left(  \delta \bar \Psi_0^{(0)} +  \delta \bar \Phi_{\parallel 0} \right)  + \sum_{\ell \in \mathbb Z} \frac{e^{i \ell \vartheta_0}}{\tau} \int {\cal E} d {\cal E} d \lambda
\nonumber  \\ & & \times \sum_\sigma \left\{  \left[ 1 - \frac{\omega_{*n}}{\omega} - \frac{\omega_{*T}}{\omega} \left( \frac{m {\cal E}}{T_0} - \frac{3}{2} \right) \right] \frac{\bar F_0}{n_0}
 \frac{\omega^2 \tau_b}{\Lambda^2 \dot{\bar\theta}_c^2 - \omega^2}\right\}_e \overline{\left(\frac{e^{- i \ell \vartheta_0}}{{\cal J} \bar B_0}\right)} \, \overline{\left( \frac{|\nabla r|}{\hat \kappa_\perp} \delta \bar \Phi_{\parallel 0}  \right)_{\vartheta_0}} 
  =  0 \;\; . \nonumber
\end{eqnarray}
This equation allows solving for $\delta \bar \Phi_{\parallel 0}$ as a function of $\delta \bar \Psi_0^{(0)}$ and shows that nonconstant Fourier harmonics in $\delta \bar \Phi_{\parallel 0}$ are predominantly due to $\vartheta_0$ scale modulation in the plasma equilibrium magnetic field $B_0$.
Equation (\ref{eq:qnKAWlim}) also reduces exactly to the well known result of a uniform plasma slab \cite{chen91,chen21}, where KAW absorption is dominated by electron Landau damping. In the present case, it is evident that Landau damping is due to circulating electrons, while trapped
electron contribution due to precession resonance is suppressed here, due to the considered frequency ordering.
Furthermore, the structure of the resonant denominator suggests that electron Landau damping is characterized by sharply localized radial structures corresponding to $|\Lambda| = |q R_0 k_\parallel| \rightarrow 0$.
 Exponentially small ion Landau damping is also readily restored by perturbatively adding its contribution on the right hand side of
Eq. (\ref{eq:qnKAWlim}), which becomes
\begin{eqnarray}
 & &  \hskip - 2em \left( \frac{1}{\tau} + \frac{\omega_{*ni}}{\omega} \right) \frac{|\nabla r|}{\hat \kappa_\perp} \delta \bar \Phi_{\parallel 0} +  \left[ \left(1 - \frac{\omega_{*ni}}{\omega} \right) \left(
1-\Gamma_0(b) \right) \right.  \label{eq:qnKAWlimion} \\ & & \left. - \frac{\omega_{*Ti}}{\omega} b \left( \Gamma_0(b) - \Gamma_1(b)\right) \right]  \frac{|\nabla r|}{\hat \kappa_\perp} \left(  \delta \bar \Psi_0^{(0)} +  \delta \bar \Phi_{\parallel 0} \right)  + \sum_{\ell \in \mathbb Z} \frac{e^{i \ell \vartheta_0}}{\tau} \int {\cal E} d {\cal E} d \lambda
\nonumber  \\ & & \times \sum_\sigma \left\{  \left[ 1 - \frac{\omega_{*n}}{\omega} - \frac{\omega_{*T}}{\omega} \left( \frac{m {\cal E}}{T_0} - \frac{3}{2} \right) \right] \frac{\bar F_0}{n_0}
 \frac{\omega^2 \tau_b}{\Lambda^2 \dot{\bar\theta}_c^2 - \omega^2}\right\}_e \overline{\left(\frac{e^{- i \ell \vartheta_0}}{{\cal J} \bar B_0}\right)} \, \overline{\left( \frac{|\nabla r|}{\hat \kappa_\perp} \delta \bar \Phi_{\parallel 0}  \right)_{\vartheta_0}} 
 \nonumber \\ & & \hskip - 1em = - \sum_{\ell \in \mathbb Z} e^{i \ell \vartheta_0}  \int {\cal E} d {\cal E} d \lambda \sum_\sigma \left\{  \left[ 1 - \frac{\omega_{*n}}{\omega} - \frac{\omega_{*T}}{\omega} \left( \frac{m {\cal E}}{T_0} - \frac{3}{2} \right) \right] \frac{\bar F_0}{n_0} \sum_{p \in \mathbb Z} \frac{p^2 \omega_b^2 \tau_b}{p^2 \omega_b^2 - \omega^2}
 \right. \nonumber \\
 & &  \left.  \times  e^{i \Lambda {\rm sgn}(\vartheta_1) \bar \theta_c} \overline{\left(\frac{e^{i p \theta_c- i \ell \vartheta_0}}{{\cal J} \bar B_0} J_0 e^{-i \hat Q_B}  \right)} \, \overline{e^{i \hat Q_B}  J_0 \left( \frac{|\bm \nabla r|}{\hat \kappa_\perp} \delta \bar \Phi_{\parallel 0} \right) e^{-i \Lambda {\rm sgn}(\vartheta_1) \bar \theta_c -ip\theta_c}}  \right\}_i \;\; . \nonumber
\end{eqnarray}
Note that, here, consistent with the perturbation approach, the ion Landau damping on the right hand side is evaluated assuming $\delta \bar \Phi_{\parallel 0}$ as obtained from the solution of Eq. (\ref{eq:qnKAWlim}) without electron Landau damping contribution. Meanwhile, noting the present frequency ordering and $b \sim {\cal O}(1)$, we have dropped a ${\cal O}(\omega_d/\omega)$ term for simplicity, which has a similar form as the term on the right hand side and could be readily restored if needed. 
Given the relation between $\delta \bar \Phi_{\parallel 0}$ and $\delta \bar \Psi_0^{(0)}$ as obtained from the solution of Eq. (\ref{eq:qnKAWlim}) or Eq. (\ref{eq:qnKAWlimion}), 
the KAW mode structure and dispersive properties are obtained by solving the vorticity equation, Eq. (\ref{eq:gkvortKAW}), which, noting Eqs. (\ref{eq:dKfluidion}) and (\ref{eq:dKeleave}), can be cast as
\begin{equation} \partial_{\vartheta_1}^2 \delta \bar \Psi_{0}^{(0)} + \overline{\left\{\frac{\omega^2 {\cal J}^{2} B_{0}^{2}}{v_A^2} \left[ \left(1 - \frac{\omega_{*ni}}{\omega} \right) 
\frac{1-\Gamma_0(b)}{b}  - \frac{\omega_{*Ti}}{\omega} \left( \Gamma_0(b) - \Gamma_1(b)\right) \right] \left( \delta \bar \Psi_0^{(0)} + \delta \bar \Phi_{\parallel 0}\right) \right\}_{\vartheta_0}} 
= 0 \; . \label{eq:gkvortKAWlim}
\end{equation}

\subsection{KAW propagation and absorption in the long wavelength limit}
\vskip 1em

The complex features of KAW propagation and absorption in general tokamak geometry can be made more explicit in the long wavelength limit, where Eqs. (\ref{eq:gkvortKAW}) and (\ref{eq:qnKAW}) can be solved by perturbation expansion. 
Detailed derivations are given in Appendix A, while only the important physics are highlighted here. 

For simplicity, let us consider the case of equilibrium distribution functions that are symmetric in $v_\parallel$ and
up-down symmetric magnetic equilibria, although the methodology presented in Appendix A is applicable in general.
Following Ref. \cite{zonca98}, in Appendix A it is shown that the long wavelength KAW spectrum near the accumulation point $\Lambda^2 \rightarrow 0$
is obtained for $\vartheta_1 = {\cal O}(\omega_A^{3/2}/\omega^{3/2})$; \idest, $\Lambda^2 = {\cal O}(\omega^3/\omega_A^3)$.
The small FLR and FOW asymptotic expansions are carried out in the smallness parameter $\omega_d/\omega_b$ 
and are obtained for $b = {\cal O}(\omega/\omega_A)$ and $\omega_d^2/\omega_b^2 = {\cal O}(\omega/\omega_A)$.
The gyrokinetic particle compression response, thus, is formally written as 
\begin{equation}
\delta \hat K = \delta \hat K^{(0)} +  \delta \hat K^{(1)} +  \delta \hat K^{(2)} +  \delta \hat K^{(3)} + ... \; ; \label{eq:dKasympt}
\end{equation} 
and similar formal expansions are assumed for the field variables. Equations (\ref{eq:gkvortKAW}) and (\ref{eq:qnKAW}) are then  systematically solved order by order, showing that the
first two orders reproduce the results discussed already in Section 3. The next two orders of Eq. (\ref{eq:qnKAW}) provide explicit expressions of $\delta \bar \Phi_{\parallel 0}^{(2)}$ and $\delta \bar \Phi_{\parallel 0}^{(3)}$
in terms of $\delta \bar \Psi_{0}^{(0)}$, while Eq. (\ref{eq:gkvortKAW}) yields $\delta \bar \Psi_{0}^{(1)} = \delta \bar \Psi_{0}^{(2)} = \delta \bar \Psi_{0}^{(3)} = 0$ and, thus, all pure $\vartheta_1$ 
dependences can be incorporated in $\delta \bar \Psi_{0}^{(0)}$. Meanwhile, the ${\cal O}(\omega^3/\omega_A^3)$ $\vartheta_0$-averaged
gyrokinetic vorticity equation yields Eq. (\ref{eq:gkvortKAW6}), reproduced here for the readers' convenience:
\begin{eqnarray} & & \hspace*{-2em} \partial_{\vartheta_1}^2 \delta \bar \Psi_{0}^{(0)} + \Lambda^2 \delta \bar \Psi_{0}^{(0)} -
\overline{\left\{\frac{\omega^2 {\cal J}^{2} B_{0}^{2}}{v_A^2} \left[ \frac{3}{4} b \left(1 - \frac{\omega_{*pi}}{\omega} - \frac{\omega_{*Ti}}{\omega} \right) 
\right] \left( \delta \bar \Psi_0^{(0)} + \delta \bar \Phi_{\parallel 0}^{(0)}\right) \right\}_{\vartheta_0}} 
\nonumber \\ & & + \overline{\left[\frac{\omega^2 {\cal J}^{2} B_{0}^{2}}{v_A^2} \left(1 - \frac{\omega_{*pi}}{\omega} \right) 
\delta \bar \Phi_{\parallel 0}^{(2)} \right]_{\vartheta_0}} -  \sum  \int {\cal E} d {\cal E} d \lambda  \sum_\sigma \tau_b \overline{\left[\frac{4\pi  \omega {\cal J} B_0^2}{k_\vartheta^2 c^2\bar B_0} 
\frac{e  \omega_d}{\hat \kappa_\perp |\bm \nabla r|}  \frac{\delta \left(  \delta \hat K^{(1)} \right)}{\delta \left(  \delta \bar \Phi_{\parallel 0}^{(0)} \right)}  \delta \bar \Phi_{\parallel 0}^{(2)}\ \right]}  \nonumber \\ & & \hspace*{2em}
-  \sum  \int {\cal E} d {\cal E} d \lambda  \sum_\sigma \tau_b \overline{\left[\frac{4\pi  \omega {\cal J} B_0^2/\bar B_0}{k_\vartheta^2 c^2} 
e \frac{\omega_d}{\hat \kappa_\perp |\bm \nabla r|}  \left( \delta \hat K^{(3)} -  \frac{k_\vartheta^2 \hat \kappa_\perp^2 v_\perp^2}{4 \Omega^2} \delta \hat K^{(1)}  \right)\right]}  \label{eq:gkvortKAW6long}  
= 0 \; . 
\end{eqnarray}
Here, $\delta \left(  \delta \hat K^{(1)} \right)/\delta \left(  \delta \bar \Phi_{\parallel 0}^{(0)} \right)$ denotes the functional derivative of $\delta \hat K^{(1)}$ with respect to $\delta \bar \Phi_{\parallel 0}^{(0)}$ \cite{gelfand63}.

Following Appendix A, it is possible to show that Eq. (\ref{eq:gkvortKAW6long}) can be cast in the form of Eq. (\ref{eq:gkvortKAWlongapp}), which is also reproduced below for the readers' convenience; that is,
\begin{equation}
 \partial_{\vartheta_1}^2 \delta \bar \Psi_{0}^{(0)} + \Lambda^2 \delta \bar \Psi_{0}^{(0)} - s^2 \vartheta_1^2 k_\vartheta^2 \bar \rho_{Li}^2 Q^2 \delta \bar \Psi_{0}^{(0)} = 0 \; , \label{eq:gkvortKAWlong}
\end{equation}
with $\bar \rho_{Li}^2 = T_{0i}/(m_i \bar\Omega_i^2)$ and $\bar \Omega_i = e \bar B_0/(m_i c)$. Expressions of $\Lambda^2$ and $Q^2$, given in Appendix A for arbitrary tokamak geometry and particle distribution functions in the 
form of quadratures involving equilibrium quantities only, generalize earlier results given for circular geometry and simplified particle responses \cite{zonca96b,chavdarovski09,zonca10,chavdarovski14,zonca98}. 
Following Ref. \cite{zonca98}, Appendix B also provides a detailed derivation of the $\Lambda^2$ and $Q^2$ functions for circular magnetic flux surfaces and well-circulating particles, showing that the present general 
approach reproduces known results in the relevant limits. In particular, it can be shown that the well-known KAW dispersion relation in nonuniform plasmas is obtained for $\omega_{bi}/\omega \rightarrow 0$, which
yields
\begin{eqnarray}
\Lambda^2 & = &  \frac{\omega^2}{\omega_A^2}  \left( 1 - \frac{\omega_{*pi}}{\omega} \right) \; , \label{eq:llkbm} \\
Q^2  & = & \frac{\omega^2}{\omega_A^2} \left[ \frac{3}{4} \left( 1 - \frac{\omega_{*pi}}{\omega} - \frac{\omega_{*Ti}}{\omega} \right) + \frac{\left( 1 - \omega_{*pi}/\omega \right)^2}{(1/\tau+ \omega_{*ni}/\omega)}  
\right] \label{eq:qqkbm} \; .
\end{eqnarray}
Equations (\ref{eq:llkbm}) and (\ref{eq:qqkbm}), as expected, can also be obtained from Eqs. (\ref{eq:qnKAWlim}) and  (\ref{eq:gkvortKAWlim}) for $b\ll 1$ and dropping electron Landau damping.

The structure of Eq. (\ref{eq:gkvortKAWlong}) with vanishing boundary conditions for  $\delta \bar \Psi_{0}^{(0)}$ as $^2 \vartheta_1^2 \rightarrow \infty$ is of a Weber equation, which admits an infinite spectrum of
solutions written in terms of the parabolic cylinder function $U(a,x)$ that solves the standard differential equation $U''-(x^2/4+a)U=0$ with $\sim e^{-x^2/4} x^{-a-1/2}$ asymptotic behavior for $x\rightarrow \infty$.
One can show \cite{zonca14a,zonca14b}
\begin{equation}
\lim_{\vartheta_1 \rightarrow 0^+} \frac{\partial_{\vartheta_1} \delta \bar \Psi_{0}^{(0)}}{\delta \bar \Psi_{0}^{(0)}} = - 2 Q^{1/2} |s k_\vartheta \bar \rho_{Li}|^{1/2} \frac{\Gamma(3/4-\Lambda^2/4  |s k_\vartheta \bar \rho_{Li}| Q)}{\Gamma(1/4-\Lambda^2/4  |s k_\vartheta \bar \rho_{Li}| Q)} \; , \label{eq:logder}
\end{equation}
where $\Gamma(x)$ denotes the Euler Gamma-function. Assuming solutions in the form $\delta \bar \Psi_{0}^{(0)} \sim e^{i \Lambda |\vartheta_1|}$, consistent with Eq. (\ref{eq:GFLDR}) as $|\vartheta_1| \rightarrow 0$, it can be verified that
Eq. (\ref{eq:logder}) recovers $i \Lambda$ as logarithmic derivative in the limit $\vartheta_1 \rightarrow 0^+$ for $|\Lambda^2|/4 s^2 k_\vartheta^2 \bar \rho_{Li}^2 Q^2 \gg 1$. Equation (\ref{eq:logder})  can then be used to extend the GFLDR to short wavelengths, including KAW \cite{zonca14a,zonca14b,zonca96a}
\begin{equation}
- 2 Q^{1/2} |s k_\vartheta \bar \rho_{Li}|^{1/2} \frac{\Gamma(3/4-\Lambda^2/4 |s k_\vartheta \bar \rho_{Li}| Q)}{\Gamma(1/4-\Lambda^2/4 |s k_\vartheta \bar \rho_{Li}| Q)} = \delta \bar W_f + \delta \bar W_k \; . \label{eq:KAWGFLDR}
\end{equation}
Even and odd solutions corresponding to the discretized KAW spectrum are obtained for 
\begin{eqnarray}
\frac{\Lambda^2}{4 |s k_\vartheta \bar \rho_{Li}| Q} & = & \ell + \frac{1}{4} \;\;\;\;\; \ell \in \mathbb N  \;\;\;\;\;  ({\rm even}) \; , \label{eq:KAWdiscrete} \\
\frac{\Lambda^2}{4 |s k_\vartheta \bar \rho_{Li}| Q} & = & \ell + \frac{3}{4} \;\;\;\;\; \ell \in \mathbb N  \;\;\;\;\;  ({\rm odd})  \nonumber \; . 
\end{eqnarray}
Near $k_\parallel qR_0 = 0$, Eq. (\ref{eq:KAWdiscrete}) suggest that time asymptotic solutions should be characterized by $k_r^2 \bar \rho_{Li}^2 = 4 |s k_\vartheta \bar \rho_{Li}| (\ell +1/4)/Q$ (even modes) and 
$k_r^2 \bar \rho_{Li}^2 = 4 |s k_\vartheta \bar \rho_{Li}|(\ell +3/4)/Q$ (odd modes), consistent with the WKB dispersion relation. In either case, $|k_r/k_\vartheta| \sim O(|k_\vartheta \bar \rho_{Li}|^{-1/2}) \gg 1$, consistent with the present asymptotic approach.
Equation (\ref{eq:KAWdiscrete}) does not represent physical solutions of the real system, which are given by Eq. (\ref{eq:KAWGFLDR}). Nonetheless, the even/odd roots of Eq. (\ref{eq:KAWdiscrete}) are a useful visualization of the discrete KAW spectrum near the 
$k_\parallel qR_0 = 0$ accumulation point. Meanwhile, the frequency difference between neighboring roots is 
\begin{equation}
\Delta \omega \simeq 4 |s k_\vartheta \bar \rho_{Li}| Q \left( \frac{\partial \Lambda^2}{\partial \omega} \right)^{-1} \; . \label{eq:deltaomega}
\end{equation}
This means that, in an initial value problem, it takes  a finite time for the distinctive features of neighboring roots to manifests themselves; and that the most likely observed behavior would be the superposition of multiple KAW in a wavepacket due, \exgra, to finite growth time and/or nonlinear characteristic time scale. This remark is crucial to understand the intrinsic time scales involved with the discretization of the kinetic continuous spectrum and completes the discussion anticipated in Section 3 about the fundamental spatial scales connected with the concept of kinetic continuous spectrum itself. 

As a final remark, it is worthwhile noting that Eqs. (\ref{eq:gkvortKAW}) and (\ref{eq:qnKAW}), and their long wavelength limit Eq. (\ref{eq:gkvortKAWlong}), do not describe KAW only. More generally, they describe the role of FLR/FOW effects on the  
discretization of kinetic continuous spectra  with Alfv\'en ic, acoustic, or mixed polarization, which is determined self-consistently with the governing equations. For example, Eq. (\ref{eq:KAWGFLDR}) describes the formation of kinetic ion acoustic waves near the BAAE accumulation point shown in Fig. \ref{fig3}. Recalling that the present analysis provides an accurate description of an asymptotic limit of the local plasma response, determining whether or not local fluctuation structures manifest themselves as physical fluctuations in cases of practical interest (see, \exgra, \cite{heidbrink25,ma23,xdu24}) also depends on the meso- and macro-scale physics, which is embedded in $\delta \bar W_f + \delta \bar W_k$ in the GFLDR.

\section{Discussions and conclusions}
\vskip 1em

In this work, we have developed a general linear gyrokinetic theory of the low-frequency shear Alfv\'en 
wave and slow sound wave continuous spectra in toroidal plasmas, accounting self-consistently for kinetic particle compressibility, 
finite parallel electric field effects and resonant wave–particle interactions. The formulation applies to general axisymmetric magnetic equilibria and 
equilibrium particle distribution functions within the adopted orderings and naturally extends to the description of dispersive wave propagation and collisionless absorption when 
finite Larmor radius and magnetic drift orbit width effects are retained. 
A distinctive feature of the present approach is that fluctuation polarization is not prescribed a priori, but emerges self-consistently from the coupled 
quasineutrality and vorticity equations, both in the kinetic-continuum limit and when dispersive effects are accounted for.

A fundamental issue addressed in this work is the physical meaning of the continuous spectrum within the framework of gyrokinetic theory. 
In fact, a peculiar feature of toroidal plasmas, such as tokamaks, with respect to simpler magnetic configurations, such as a sheared slab or an axisymmetric dipole magnetic field, 
is the finite geodesic curvature of equilibrium magnetic field lines. The associated magnetic drifts introduce particle motion across magnetic flux surfaces, 
which themselves provide the spatial label of the continuous spectrum. Thus, the notion of continuous spectrum, conventionally associated with the limit of vanishing radial wavelength, needs to be reconsidered 
when accounting for the microscopic dynamics of the plasma. As shown in this work, the kinetic continuous spectrum should be rigorously understood as an intermediate asymptotic limit of the local plasma response, 
characterized by radial spatial scales that are sufficiently short with respect to the poloidal wavelength of the considered fluctuation and sufficiently long with respect to the characteristic particle gyroradii and magnetic drift orbit widths. 
Nonetheless, even in this limit, geodesic curvature effects remain crucial in determining the kinetic compressional response and, thereby, the dispersion and polarization of the underlying fluctuations. In fact, 
the increasing magnetic drift frequency with decreasing radial wavelength compensates for the corresponding decrease of the geometrical factor multiplying the shear Alfv\'en stream function in the gyrokinetic equation, 
such that the magnetic drift contribution to the kinetic response remains finite in the adopted asymptotic limit. This observation illuminates the fundamentally different nature of the kinetic continuous spectrum in tokamak plasmas 
with respect to magnetic configurations where particle magnetic drifts are confined to magnetic flux surfaces.

The intrinsic multiscale character of low-frequency fluctuations in toroidal plasmas allows a systematic separation of the spatial scales associated with equilibrium geometry modulations 
and those characterizing the flute-like nature of the underlying parallel mode structures. Meanwhile, the kinetic particle response introduces spatial and temporal nonlocality associated with particle equilibrium motion 
and resonant wave–particle interactions. As a consequence, although the periodicity of the equilibrium allows a Bloch-Floquet representation of the fluctuation structures, the conventional construction of Floquet solutions 
based on a finite-dimensional monodromy matrix is generally not applicable. The extended Bloch-Floquet formulation developed here overcomes this difficulty and yields the renormalized plasma response 
in quadrature form. In particular, the Floquet exponent enters explicitly the resonance condition for circulating particles, while being itself determined by the resonant particle response. 
Thus, the present formulation provides a self-consistent description of the interplay between fluctuation parallel mode structures, polarization and resonant wave–particle interactions, 
without requiring restrictive assumptions on equilibrium geometry or particle dynamics beyond the adopted gyrokinetic and asymptotic orderings. 
The generality of this formulation also provides a systematic basis for assessing the validity of approximations commonly adopted in the description of low-frequency fluctuations in toroidal plasmas.

An important consequence of the present analysis is the clarification of the role of fluctuation polarization in determining the stability properties of low-frequency drift Alfv\'en waves. 
In this respect, the distinction between predominantly Alfv\'enic and acoustic fluctuations is not merely a characterization of their electromagnetic properties, but reflects different susceptibilities 
of the plasma to resonant excitation. As discussed within the framework of the general fishbone-like dispersion relation, the preferential excitation of Alfv\'enic polarization by energetic particles 
is connected with two distinct physical mechanisms. On the one hand, predominantly Alfv\'enic fluctuations are generally characterized, in the cases investigated here, by significantly weaker collisionless 
damping than their predominantly acoustic counterparts. On the other hand, the different dispersive properties of the underlying branches, reflected by the frequency dependence of the generalized inertia, 
imply a stronger response of Alfv\'enic fluctuations to a given potential energy contribution associated with resonant energetic particles. 
The latter feature is particularly important, since the relative ease with which different branches of the kinetic continuous spectrum can be resonantly excited cannot be inferred from their damping rates alone. 
These considerations also emphasize the importance of experimentally assessing fluctuation polarization, in addition to fluctuation frequency, for the proper identification of low-frequency 
drift Alfv\'en waves in toroidal plasmas and the interpretation of their excitation mechanisms.

Retaining finite Larmor radius and magnetic drift orbit width effects allows us to overcome the conceptual abstraction involved in the definition of the kinetic continuous spectrum and to address 
fluctuation propagation and absorption on physical spatial and temporal scales. Although the discussion in Section 4 has been primarily focused on kinetic Alfv\'en waves, the present formulation 
is not restricted to fluctuations of Alfv\'enic polarization. Rather, it provides a general framework for describing the dispersive extensions of the different branches of the kinetic continuous spectrum, 
with their peculiar polarization properties determined self-consistently by the governing equations. In this respect, the kinetic continuous spectrum and the corresponding dispersive waves can be 
regarded as different manifestations of the same underlying plasma response. Their polarization properties, which play a crucial role in determining the susceptibility to resonant excitation, 
are also important in determining their propagation and collisionless absorption. Thus, the self-consistent determination of fluctuation polarization is central to the description of both the 
excitation of fluctuations and the subsequent spatial redistribution and absorption of their energy.

A peculiar feature of the low-frequency plasma response in toroidal geometry is that finite parallel electric fields are not solely connected with finite Larmor radius effects, 
but also with the kinetic compressional particle response associated with magnetic drifts. The relative importance of the latter increases with decreasing fluctuation frequency, 
such that the properties of low-frequency kinetic Alfv\'en waves in toroidal plasmas cannot generally be inferred by extrapolating their well-known high-frequency behavior. 
While the present formulation recovers the conventional kinetic Alfv\'en wave dispersion and damping in the appropriate high-frequency limit, their low-frequency propagation and 
collisionless absorption reflect the peculiarities of particle equilibrium motion, fluctuation polarization and resonant wave–particle interactions in toroidal geometry. 
Besides illuminating the physical mechanisms underlying these processes, the analytical description developed here provides a stringent benchmark for 
gyrokinetic calculations of wave propagation and absorption in toroidal plasmas.

Although the present work is concerned with linear plasma response, its implications extend to the description of nonlinear fluctuation dynamics. 
In fact, the spectral properties of the underlying fluctuations, including their polarization, dispersion, propagation and collisionless absorption, 
are crucial ingredients in determining nonlinear wave-wave and wave-particle interactions. This is particularly relevant in the low-frequency range 
considered here, where resonantly excited drift Alfv\'en waves may play an important role in mediating cross-scale interactions between energetic particles, 
Alfv\'enic fluctuations and drift wave turbulence. Furthermore, the generation of zonal electromagnetic fields and their impact on turbulent transport 
depend sensitively on the spectral properties of the fluctuations involved and on the spatial scales of their propagation and absorption. 
Thus, the present formulation provides a theoretical foundation for investigating these processes on the basis of a self-consistent description of 
the underlying kinetic plasma response, with potentially important implications for the understanding of energetic particle effects on turbulent transport in burning plasmas.

Finally, we note that, although the present analysis has been developed for axisymmetric toroidal plasmas, 
its underlying physical and mathematical structure suggests possible extensions to non-axisymmetric magnetic configurations, such as stellarators. 
Such extensions require proper consideration of the peculiarities of particle equilibrium motion and of the spatial structure of fluctuations in the absence of toroidal symmetry. 
More generally, the present work illustrates how the interplay between magnetic geometry, microscopic particle dynamics and the multiscale nature of fluctuations determines the spectral properties of magnetized plasmas. 
The unified description of kinetic continuous spectra and their dispersive extensions developed here provides a natural framework for addressing this interplay 
and its consequences for the polarization, excitation, propagation and absorption of fluctuations in toroidal plasmas.

\appendix

\section{Kinetic Alfv\'en Waves in the long wavelength limit}
\vskip 1em

Let us consider Eqs. (\ref{eq:gkvortKAW}) and (\ref{eq:qnKAW}), derived in Sec. 4, and specialize them to the case of small FLR/FOW. In Ref. \cite{zonca98}, it has been shown that the
self consistent ordering for KAW perturbation expansion near the accumulation point of the continuous spectrum is obtained considering $\Lambda^2 = {\cal O}(\omega^3/\omega_A^3)$ and,
thus, $\vartheta_1 = {\cal O}(\omega_A^{3/2}/\omega^{3/2})$, which is consistent with $\Lambda^2 \rightarrow 0$ since, formally, $\Lambda^2 = {\cal O}(\omega^2/\omega_A^2)$ as shown in Sec. 2.
For consistency, small FLR and FOW asymptotic expansions in the smallness parameter $\omega_d/\omega_b$ are obtained for $b = {\cal O}(\omega/\omega_A)$ and $\omega_d^2/\omega_b^2 = {\cal O}(\omega/\omega_A)$.
Thus, $\delta \hat K = \delta \hat K^{(0)} +  \delta \hat K^{(1)} +  \delta \hat K^{(2)} +  \delta \hat K^{(3)} + ...$ denotes the asymptotic expansion of the gyrokinetic particle compression response and
similar formal expansions are used for the field variables. Furthermore, we limit the present analysis to equilibrium distribution functions that are symmetric in $v_\parallel$ and to 
up-down symmetric equilibria in order to avoid complications discussed in Sec. 3. This does not limit the generality of the present approach, 
which can be adopted also for generic $\bar F_0$ and up-down asymmetric equilibria with additional formal complications.

By inspection of the vorticity equation, Eq. (\ref{eq:gkvortKAW}), we recognize that both inertia and pressure-curvature coupling terms due to kinetic particle compression responses are formally of ${\cal O}(\omega^2/\omega_A^2)$,
that is of fourth order in our asymptotic expansion. Thus, since we formally have 
$$\partial_\vartheta^2 = \partial_{\vartheta_0}^2 + 2 \partial_{\vartheta_0} \partial_{\vartheta_1} + \partial_{\vartheta_1}^2 \; ,$$
we readily have 
$$\partial_{\vartheta_0}^2 \delta \bar \Psi_{0}^{(0)} = \partial_{\vartheta_0}^2 \delta \bar \Psi_{0}^{(1)} = \partial_{\vartheta_0}^2 \delta \bar \Psi_{0}^{(2)} = \partial_{\vartheta_0}^2 \delta \bar \Psi_{0}^{(3)}  = 0 \; ;$$
and, thus, $\delta \bar \Psi_{0}^{(1)} = \delta \bar \Psi_{0}^{(2)} = \delta \bar \Psi_{0}^{(3)} = 0$, since all pure $\vartheta_1$ dependences can be incorporated in $\delta \bar \Psi_{0}^{(0)}$ \cite{zonca98}.
Meanwhile, the lowest order solution for the gyrokinetic particle compression response yields
\begin{equation}
\delta \hat K^{(0)} =  \frac{e}{m} \frac{Q\bar F_0}{\omega} \frac{ \omega (|\bm \nabla r| \delta \bar \Phi_{\parallel 0}^{(0)}/\hat \kappa_\perp) + n \bar{\omega}_d (|\bm \nabla r| \delta \bar \Psi_{0}^{(0)}/\hat \kappa_\perp)}
{n\bar\omega_{d} + \Lambda \dot{\bar\theta}_c {\rm sgn}(\vartheta_1) - \omega} \; ,
\label{eq:fdklow0}
\end{equation} 
which, substituted back into the quasineutrality condition, Eq. (\ref{eq:quasineutrality}), yields Eq. (\ref{eq:qnsolvel0}) with $\delta \bar \Phi_{\parallel 0} \rightarrow \delta \bar \Phi_{\parallel 0}^{(0)}$.
At next order, the gyrokinetic particle compression response is given by
\begin{eqnarray}
\hspace{-0em} \delta \hat K^{(1)} & = & \sum_{p \neq 0}  \left\{\frac{e}{m} \frac{Q\bar F_0}{\omega}  \overline{\frac{ \omega (|\bm \nabla r| \delta \bar \Phi_{\parallel 0}^{(1)}/\hat \kappa_\perp) e^{-i \Lambda {\rm sgn}(\vartheta_1) \bar \theta_c 
 -ip\theta_c}}{n\bar\omega_{d} + p \omega_b + \Lambda \dot{\bar\theta}_c {\rm sgn}(\vartheta_1) - \omega}  }
 \right. \label{eq:fdklow1}\\
 & & \hspace{-4em} +\frac{e}{m}  \frac{Q\bar F_0}{\omega} \frac{\overline{\omega_d e^{
 -ip\theta_c}}}{n\bar\omega_{d} + p \omega_b + \Lambda \dot{\bar\theta}_c {\rm sgn}(\vartheta_1) - \omega}  \left(|\bm \nabla r| (\delta \bar \Psi_{0}^{(0)} + \delta \bar \Phi_{\parallel 0}^{(0)} ) /\hat \kappa_\perp \right) e^{-i \Lambda {\rm sgn}(\vartheta_1) \bar \theta_c} 
 \nonumber  \\
 & & \hspace{-4em} - \frac{e}{m}  \frac{Q\bar F_0}{\omega} \frac{ n \bar\omega_d \left(\overline{\omega_d e^{
 -ip\theta_c}}\right) }{n\bar\omega_{d} + p \omega_b + \Lambda \dot{\bar\theta}_c {\rm sgn}(\vartheta_1) - \omega}  \frac{\left(|\bm \nabla r| (\delta \bar \Psi_{0}^{(0)} + \delta \bar \Phi_{\parallel 0}^{(0)} ) /\hat \kappa_\perp \right)}{n\bar\omega_{d}
 + \Lambda \dot{\bar\theta}_c {\rm sgn}(\vartheta_1) - \omega} e^{-i \Lambda {\rm sgn}(\vartheta_1) \bar \theta_c} 
 \nonumber  \\
  & & \hspace{-4em} \left.  - \frac{e}{m}  \frac{Q\bar F_0}{\omega} \frac{ \Lambda \dot{\bar\theta}_c {\rm sgn}(\vartheta_1)  \left(\overline{\omega_d e^{
 -ip\theta_c}}\right) }{n\bar\omega_{d} + p \omega_b + \Lambda \dot{\bar\theta}_c {\rm sgn}(\vartheta_1) - \omega}  \frac{\left(|\bm \nabla r| \delta \bar \Phi_{\parallel 0}^{(0)} /\hat \kappa_\perp \right)}{n\bar\omega_{d}
 + \Lambda \dot{\bar\theta}_c {\rm sgn}(\vartheta_1) - \omega} e^{-i \Lambda {\rm sgn}(\vartheta_1) \bar \theta_c}  \right\}
 e^{i \Lambda {\rm sgn}(\vartheta_1) \bar \theta_c  + i p \theta_c}  
 \;\; . \nonumber 
\end{eqnarray} 
The third and fourth line in this expression correspond to higher order contributions, respectively of ${\cal O}(\omega^{2}/\omega_A^{2})$ and ${\cal O}(\omega^{3/2}/\omega_A^{3/2})$ with respect to the 
leading order terms, due to the presence of the two resonant denominators and the asymptotic expansion approach adopted here. Thus, they will be dropped, since they contribute to higher order than that required for evaluation 
of KAW propagation and absorption. 
The corresponding simplified $\delta \hat K^{(1)}$ response can be formally written as
\begin{equation}
\delta \hat K^{(1)}  = \sum_{p \neq 0} e^{i p \theta_c} \delta \hat K^{(1)}_p \; , \label{eq:dhatkdec}
\end{equation}
with the partial amplitudes $ \delta \hat K^{(1)}_p$ defined by inspection of Eq. (\ref{eq:fdklow1}). 
Upon substitution into Eq. (\ref{eq:quasineutrality}), this yields the results of Eq. (\ref{eq:qnsolvelow}) where the nonconstant Fourier harmonics of the
parallel electric-field, $(|\bm \nabla r| \delta \bar \Phi_{\parallel 0}^{(1)}/\hat \kappa_\perp)_l$, are solved as a function of $(|\bm \nabla r| \delta \bar \Psi_{0}^{(0)}/\hat \kappa_\perp)$ with an approximated expression of the quasineutrality matrix $A_{l,l'}$, truncated to $l,l' \neq 0$,
and with $C_l$ renormalized as \cite{chavdarovski09,chavdarovski14}
\begin{equation}
C_l \mapsto \left( 1 + C_0/A_{0,0} \right) C_l \label{eq:Clrenorm}
\end{equation}
for $l\neq 0$.
When used in Eq. (\ref{eq:gkvortKAW}) neglecting FLR/FOW, Eq. (\ref{eq:fdklow1}) consistently yields the generalized inertia expression of Eq. (\ref{eq:Lambda2gen}), with repeated Fourier indexes running on  $l,l' \neq 0$ and $C_l$ given by Eq. (\ref{eq:Clrenorm}). 

At fourth order, ${\cal O}(\omega^2/\omega_A^2)$, the vorticity equation becomes
\begin{equation} \partial_{\vartheta_0}^2 \delta \bar \Psi_{0}^{(4)} +  \Lambda^2 \delta \bar \Psi_{0}^{(0)} -  \sum \sum_{\ell \neq 0} e^{i \ell \vartheta_0} \int {\cal E} d {\cal E} d \lambda  \sum_\sigma \tau_b \overline{\left[\frac{4\pi  \omega {\cal J} B_0^2/\bar B_0}{k_\vartheta^2 c^2} 
e \frac{\omega_d}{\hat \kappa_\perp |\bm \nabla r|}   e^{- i \ell \vartheta_0} \delta \hat K^{(1)} \right]}
= 0 \; . \label{eq:gkvortKAW4} 
\end{equation}
This implies that $\Lambda^2 \rightarrow 0$ at ${\cal O}(\omega^2/\omega_A^2)$, while the periodic variation of $\delta \bar \Psi_{0}^{(4)}$ in $\vartheta_0$ with zero average can be solved from Eq. (\ref{eq:gkvortKAW4}), although its explicit expression does not
modify the KAW dispersion relation as shown below. Proceeding further at fifth order, we have
\begin{eqnarray} & & \partial_{\vartheta_0}^2 \delta \bar \Psi_{0}^{(5)} + \Lambda^2 \delta \bar \Psi_{0}^{(0)} + \sum_{\ell \in \mathbb Z} e^{i \ell \vartheta_0} \overline{\left[\frac{\omega^2 {\cal J}^{2} B_{0}^{2}}{v_A^2} \left[ \left(1 - \frac{\omega_{*pi}}{\omega} \right) e^{-i \ell \vartheta_0} 
\delta \bar \Phi_{\parallel 0}^{(1)}\right) \right]_{\vartheta_0}} \nonumber \\ & & \hspace*{2em}
-  \sum \sum_{\ell \in \mathbb Z} e^{i \ell \vartheta_0} \int {\cal E} d {\cal E} d \lambda  \sum_\sigma \tau_b \overline{\left[\frac{4\pi  \omega {\cal J} B_0^2/\bar B_0}{k_\vartheta^2 c^2} 
e \frac{\omega_d}{\hat \kappa_\perp |\bm \nabla r|}   e^{- i \ell \vartheta_0} \delta \hat K^{(2)} \right]} \label{eq:gkvortKAW5}  
= 0 \; . 
\end{eqnarray}
The presence of non-vanishing $\ell = 0$ terms in Eq. (\ref{eq:gkvortKAW5}) is related to the symmetry properties of the considered magnetic equilibrium and to the discussion introduced 
in Sec. 3 below Eq. (\ref{eq:qnsolvel0}). There, we argued that breaking up-down
symmetry can produce a finite parallel electric field through geodesic-curvature effects due to $\delta \bar \Phi_{\parallel 0}^{(1)}$ and modify the corresponding KAW propagation and absorption properties by redefinition of $\Lambda^2$ at 
${\cal O}(\omega^{5/2}/\omega_A^{5/2})$. Similar argument applies to the $\ell = 0$ term due to the gyrokinetic particle compression response $\sim \delta \hat K^{(2)}$. In an up-down symmetric equilibrium with
equilibrium particle distribution functions that are symmetric in $v_\parallel$, both these terms vanish. 
Thus, similar to Eq. (\ref{eq:gkvortKAW4}), Eq. (\ref{eq:gkvortKAW5}) implies that $\Lambda^2 \rightarrow 0$ at ${\cal O}(\omega^{5/2}/\omega_A^{5/2})$, while the $\ell \neq 0$ terms yield the solution for $\delta \bar \Psi_{0}^{(5)}$ with zero average,
which, analogously to $\delta \bar \Psi_{0}^{(4)}$, does not modify the KAW dispersion relation; so, it will be left undetermined. The KAW dispersion relation is determined at sixth order, ${\cal O}(\omega^3/\omega_A^3)$, by the $\vartheta_0$ averaged
gyrokinetic vorticity equation, which annihilates $\delta \bar \Psi_{0}^{(6)}$ and yields
\begin{eqnarray} & & \hspace*{-2em} \partial_{\vartheta_1}^2 \delta \bar \Psi_{0}^{(0)} + \Lambda^2 \delta \bar \Psi_{0}^{(0)} -
\overline{\left\{\frac{\omega^2 {\cal J}^{2} B_{0}^{2}}{v_A^2} \left[ \frac{3}{4} b \left(1 - \frac{\omega_{*pi}}{\omega} - \frac{\omega_{*Ti}}{\omega} \right) 
\right] \left( \delta \bar \Psi_0^{(0)} + \delta \bar \Phi_{\parallel 0}^{(0)}\right) \right\}_{\vartheta_0}} 
\nonumber \\ & & + \overline{\left[\frac{\omega^2 {\cal J}^{2} B_{0}^{2}}{v_A^2} \left(1 - \frac{\omega_{*pi}}{\omega} \right) 
\delta \bar \Phi_{\parallel 0}^{(2)} \right]_{\vartheta_0}} -  \sum  \int {\cal E} d {\cal E} d \lambda  \sum_\sigma \tau_b \overline{\left[\frac{4\pi  \omega {\cal J} B_0^2}{k_\vartheta^2 c^2\bar B_0} 
\frac{e  \omega_d}{\hat \kappa_\perp |\bm \nabla r|}  \frac{\delta \left(  \delta \hat K^{(1)} \right)}{\delta \left(  \delta \bar \Phi_{\parallel 0}^{(0)} \right)}  \delta \bar \Phi_{\parallel 0}^{(2)}\ \right]}  \nonumber \\ & & \hspace*{2em}
-  \sum  \int {\cal E} d {\cal E} d \lambda  \sum_\sigma \tau_b \overline{\left[\frac{4\pi  \omega {\cal J} B_0^2/\bar B_0}{k_\vartheta^2 c^2} 
e \frac{\omega_d}{\hat \kappa_\perp |\bm \nabla r|}  \left( \delta \hat K^{(3)} -  \frac{k_\vartheta^2 \hat \kappa_\perp^2 v_\perp^2}{4 \Omega^2} \delta \hat K^{(1)}  \right)\right]}  \label{eq:gkvortKAW6}  
= 0 \; . 
\end{eqnarray}
Here, $\delta \left(  \delta \hat K^{(1)} \right)/\delta \left(  \delta \bar \Phi_{\parallel 0}^{(0)} \right)$ denotes the functional derivative of $\delta \hat K^{(1)}$ with respect to $\delta \bar \Phi_{\parallel 0}^{(0)}$ (see, \exgra, \cite{gelfand63}).
Equation (\ref{eq:gkvortKAW6}) generalizes the former result of \cite{zonca98}, obtained for circular magnetic flux surfaces and well-circulating particles, 
to general tokamak geometry\footnote{The extension to non up-down symmetric equilibria and general particle distribution functions can be obtained in the light of the discussion following 
Eq. (\ref{eq:gkvortKAW5}) above.} and particle orbits.
 
The gyrokinetic vorticy equation in the form of Eq. (\ref{eq:gkvortKAW6}) can be cast in the form of a KAW dispersion relation upon deriving expressions for the  kinetic particle compression responses, $\delta \hat K^{(1)}$ and $\delta \hat K^{(3)}$,
and of the generalized parallel potential $\delta \bar \Phi_{\parallel 0}^{(2)}$. We already demonstrated that the
first two orders in the perturbation expansion are identical to the solution discussed in Sec. 3 for the derivation of the kinetic continuous spectrum. Meanwhile,
\begin{eqnarray}
\hspace{-0em} \delta \hat K^{(2)} & = & \sum_{p \in \mathbb Z}  \left\{\frac{e}{m} \frac{Q\bar F_0}{\omega}  \overline{\frac{ \omega (|\bm \nabla r| \delta \bar \Phi_{\parallel 0}^{(2)}/\hat \kappa_\perp) e^{-i \Lambda {\rm sgn}(\vartheta_1) \bar \theta_c 
 -ip\theta_c}}{n\bar\omega_{d} + p \omega_b + \Lambda \dot{\bar\theta}_c {\rm sgn}(\vartheta_1) - \omega}  }
 \right.  \label{eq:fdklow2} \\
  & & -  \frac{e}{m} \frac{Q\bar F_0}{\omega}  \overline{\left(\frac{k_\vartheta^2 \hat \kappa_\perp^2 v_\perp^2}{4 \Omega^2} \right) \frac{ \omega (|\bm \nabla r| \delta \bar \Phi_{\parallel 0}^{(0)}/\hat \kappa_\perp) e^{-i \Lambda {\rm sgn}(\vartheta_1) \bar \theta_c 
 -ip\theta_c}}{n\bar\omega_{d} + p \omega_b + \Lambda \dot{\bar\theta}_c {\rm sgn}(\vartheta_1) - \omega}  }
\nonumber \\
  & & - \left.  \sum_{p' \neq p \in \mathbb Z}  \frac{\overline{\omega_d e^{-i p' \theta_c}} \delta \hat K_{p-p'}^{(1)} e^{-i \Lambda {\rm sgn}(\vartheta_1) \bar \theta_c}}{n\bar\omega_{d} + p \omega_b + \Lambda \dot{\bar\theta}_c {\rm sgn}(\vartheta_1) - \omega}  
 \nonumber  \right\}
 e^{i \Lambda {\rm sgn}(\vartheta_1) \bar \theta_c  + i p \theta_c}  
 \;\; . \nonumber 
\end{eqnarray} 
When substituted back into the quasineutrality condition, Eq. (\ref{eq:qnKAW}), this expression yields
\begin{eqnarray}
& & A_{l l'} (|\bm \nabla r| \delta \bar \Phi_{\parallel 0}^{(2)}/\hat \kappa_\perp)_{l'} = - \left( 1 - \frac{\omega_{*pi}}{\omega} \right) \overline{ \left( e^{- i l \vartheta_0}  k_\vartheta^2 \hat \kappa_\perp^2 \frac{T_{0i}/m_i}{\Omega_i^2}
\right)_{\vartheta_0}} \left( \frac{|\bm \nabla r|}{\hat \kappa_\perp} \delta \bar \Psi_{0}^{(0)} \right) \label{eq:phibar2} \\
& & \hspace*{2em} + \int {\cal E} d {\cal E} d \lambda \sum_\sigma \left\{ \tau_b \left[ 1 - \frac{\omega_{*n}}{\omega} - \frac{\omega_{*T}}{\omega} \left( \frac{m {\cal E}}{T_0} - \frac{3}{2} \right) \right] \frac{\bar F_0}{n_0}  \left[ \overline{
\left(\frac{e^{- i \ell \vartheta_0}}{{\cal J} \bar B_0} \right) \left(\frac{k_\vartheta^2 \hat \kappa_\perp^2 v_\perp^2}{4 \Omega^2} \right) 
 }  \right. \right. \nonumber  \\ & &  \hspace*{2em}  \times \frac{\omega}{n\bar\omega_{d} + \Lambda \dot{\bar\theta}_c {\rm sgn}(\vartheta_1) - \omega}   \left( \frac{|\bm \nabla r|}{\hat \kappa_\perp} \delta \bar \Phi_{\parallel 0}^{(0)} \right)  
 + \sum_{p \in \mathbb Z}  \overline{
\left(\frac{e^{i p \theta_c - i \ell \vartheta_0}}{{\cal J} \bar B_0} \right)} \, \overline{ \left(e ^{- i p \theta_c} \frac{k_\vartheta^2 \hat \kappa_\perp^2 v_\perp^2}{4 \Omega^2} \right) 
 }   \nonumber \\ & & \hspace*{2em}  \left. \left.  \times  \frac{ \omega (|\bm \nabla r| \delta \bar \Phi_{\parallel 0}^{(0)}/\hat \kappa_\perp) + n \bar{\omega}_d (|\bm \nabla r| \delta \bar \Psi_{0}^{(0)}/\hat \kappa_\perp)}
{n\bar\omega_{d} + p \omega_b + \Lambda \dot{\bar\theta}_c {\rm sgn}(\vartheta_1) - \omega} \right] \right\}_i  - \sum \sum_{p \in \mathbb Z} \int {\cal E} d {\cal E} d \lambda \sum_\sigma \tau_b \nonumber
\\ & & \hspace*{2em}  \times  \frac{T_{0i}}{n_0 e}  \sum_{p' \neq p \in \mathbb Z}  \frac{ 
\overline{\left( e^{i p \theta_c - i \ell \vartheta_0} /{\cal J} \bar B_0\right)}\, \overline{\omega_d e^{-i p' \theta_c}} 
\delta \hat K_{p-p'}^{(1)} }{n\bar\omega_{d} + p \omega_b + \Lambda \dot{\bar\theta}_c {\rm sgn}(\vartheta_1) - \omega}  \; ; \nonumber
\end{eqnarray}
for the  $l$-Fourier harmonic in $\vartheta_0$ of the generalized parallel potential, $(|\bm \nabla r| \delta \bar \Phi_{\parallel 0}^{(2)}/\hat \kappa_\perp)_l$. Equation (\ref{eq:phibar2})  
illuminates the nature of the generation of parallel electric field in toroidal geometry. The last term is due to toroidal FOW effects
associated with geodesic curvature, and it involves thermal ion as well as thermal electron responses. However, while thermal ions provide the dominant Hermitian response, connected with KAW propagation properties, 
the anti-Hermitian response defining KAW absorption is due to both thermal ions and electrons. Other terms on the right hand side, meanwhile, are genuinely due to FLR. While the first one is the usual contribution to 
parallel electric that appears in slab geometry, the second is due to normal curvature and originates from toroidal precession and magnetic drifts within the considered magnetic flux surface. This effect is important at very low
frequencies, comparable with the thermal plasma precession frequencies, and may impact on both KAW propagation and absorption. For this reason, it is not expected to be crucial for DAW excitation by EPs but rather for
shorter wavelength fluctuations for which thermal plasma effects become increasingly more important.

Proceeding further at next order, we finally have
\begin{eqnarray}
\hspace{-0em} \delta \hat K^{(3)} & = & \sum_{p \in \mathbb Z}  \left\{\frac{e}{m} \frac{Q\bar F_0}{\omega}  \overline{\frac{ \omega (|\bm \nabla r| \delta \bar \Phi_{\parallel 0}^{(3)}/\hat \kappa_\perp) e^{-i \Lambda {\rm sgn}(\vartheta_1) \bar \theta_c 
 -ip\theta_c}}{n\bar\omega_{d} + p \omega_b + \Lambda \dot{\bar\theta}_c {\rm sgn}(\vartheta_1) - \omega}  }
 \right.  \label{eq:fdklow3} \\
  & & -  \frac{e}{m} \frac{Q\bar F_0}{\omega}  \overline{\left(\frac{k_\vartheta^2 \hat \kappa_\perp^2 v_\perp^2}{4 \Omega^2} \right) \frac{ \omega (|\bm \nabla r| \delta \bar \Phi_{\parallel 0}^{(1)}/\hat \kappa_\perp) e^{-i \Lambda {\rm sgn}(\vartheta_1) \bar \theta_c 
 -ip\theta_c}}{n\bar\omega_{d} + p \omega_b + \Lambda \dot{\bar\theta}_c {\rm sgn}(\vartheta_1) - \omega}  }
\nonumber \\
   & & -  \frac{e}{m} \frac{Q\bar F_0}{\omega}  \overline{\left(\frac{k_\vartheta^2 \hat \kappa_\perp^2 v_\perp^2}{4 \Omega^2} \right) \frac{ \omega_d (|\bm \nabla r| \delta \bar \Psi_{0}^{(0)}/\hat \kappa_\perp) e^{-i \Lambda {\rm sgn}(\vartheta_1) \bar \theta_c 
 -ip\theta_c}}{n\bar\omega_{d} + p \omega_b + \Lambda \dot{\bar\theta}_c {\rm sgn}(\vartheta_1) - \omega}  }
 \nonumber\\
  & & - \left.  \sum_{p' \in \mathbb Z}  \frac{\overline{\omega_d e^{-i p' \theta_c}} \delta \hat K_{p-p'}^{(2)} e^{-i \Lambda {\rm sgn}(\vartheta_1) \bar \theta_c}}{n\bar\omega_{d} + p \omega_b + \Lambda \dot{\bar\theta}_c {\rm sgn}(\vartheta_1) - \omega}  
 \nonumber  \right\}
 e^{i \Lambda {\rm sgn}(\vartheta_1) \bar \theta_c  + i p \theta_c}  
 \;\;  , \nonumber 
\end{eqnarray} 
where we have decomposed $\delta \hat K^{(2)}$ as shown in Eq. (\ref{eq:dhatkdec}) for $\delta \hat K^{(1)}$ but including the $p=0$ harmonic. Using this expression in the quasineutrality condition, Eq. (\ref{eq:qnKAW}), 
we obtain, similar to Eq. (\ref{eq:phibar2}), 
\begin{eqnarray}
& & A_{l l'} (|\bm \nabla r| \delta \bar \Phi_{\parallel 0}^{(3)}/\hat \kappa_\perp)_{l'} =  \int {\cal E} d {\cal E} d \lambda \sum_\sigma \left\{ \tau_b \left[ 1 - \frac{\omega_{*n}}{\omega} - \frac{\omega_{*T}}{\omega} \left( \frac{m {\cal E}}{T_0} - \frac{3}{2} \right) \right] \frac{\bar F_0}{n_0} 
 \right. \label{eq:phibar3}  \\ & &  \hspace*{2em} \left.  \times \sum_{p \in \mathbb Z}  \overline{
\left(\frac{e^{i p \theta_c - i \ell \vartheta_0}}{{\cal J} \bar B_0} \right)} \, \overline{ \left(e ^{- i p \theta_c} \frac{k_\vartheta^2 \hat \kappa_\perp^2 v_\perp^2}{4 \Omega^2} \right)  
\frac{ \omega (|\bm \nabla r| \delta \bar \Phi_{\parallel 0}^{(1)}/\hat \kappa_\perp) + {\omega}_d (|\bm \nabla r| \delta \bar \Psi_{0}^{(0)}/\hat \kappa_\perp)}
{n\bar\omega_{d} + p \omega_b + \Lambda \dot{\bar\theta}_c {\rm sgn}(\vartheta_1) - \omega} } \right\}_i \nonumber \\ & & \hspace*{2em}  -  \sum \sum_{p \in \mathbb Z} \int {\cal E} d {\cal E} d \lambda \sum_\sigma \tau_b 
 \frac{T_{0i}}{n_0 e}  \sum_{p'  \in \mathbb Z}  \frac{ 
\overline{\left( e^{i p \theta_c - i \ell \vartheta_0} /{\cal J} \bar B_0\right)}\, \overline{\omega_d e^{-i p' \theta_c}} 
\delta \hat K_{p-p'}^{(2)} }{n\bar\omega_{d} + p \omega_b + \Lambda \dot{\bar\theta}_c {\rm sgn}(\vartheta_1) - \omega}  
\nonumber \\ & & \hspace*{2em}  -  \sum \sum_{p \in \mathbb Z} \int {\cal E} d {\cal E} d \lambda \sum_\sigma \tau_b 
 \frac{T_{0i}}{n_0 e}  \overline{ \left(\frac{e^{i p \theta_c - i \ell \vartheta_0}}{{\cal J} \bar B_0} \right) \left(\frac{k_\vartheta^2 \hat \kappa_\perp^2 v_\perp^2}{4 \Omega^2} \right)}  \delta \hat K_{p}^{(1)} 
\; . \nonumber
\end{eqnarray}
The physical interpretation of the various terms in Eq. (\ref{eq:phibar3})  closely follows that of Eq. (\ref{eq:phibar2}) and, thus, it will be 
omitted for brevity. Noting these results, the gyrokinetic vorticity equation, Eq. (\ref{eq:gkvortKAW6}), describing KAW propagation and absorption in the long wavelength limit becomes  \cite{zonca98}
\begin{equation}
 \partial_{\vartheta_1}^2 \delta \bar \Psi_{0}^{(0)} + \Lambda^2 \delta \bar \Psi_{0}^{(0)} - s^2 \vartheta_1^2 k_\vartheta^2 \bar \rho_{Li}^2 Q^2 \delta \bar \Psi_{0}^{(0)} = 0 \; , \label{eq:gkvortKAWlongapp}
\end{equation}
with $\bar \rho_{Li}^2 = T_{0i}/(m_i \bar\Omega_i^2)$ and $\bar \Omega_i = e \bar B_0/(m_i c)$. Here, the expression of $Q^2$ is obtained by inspection of Eqs. (\ref{eq:fdklow1})-(\ref{eq:dhatkdec}), and Eqs. (\ref{eq:fdklow2})-(\ref{eq:phibar3}),
and generalizes to arbitrary tokamak geometry the former result obtained for circular magnetic flux surfaces and well-circulating particles.

\section{The case of circular magnetic flux surfaces and well-circulating particles.}
\vskip 1em

\begin{figure}
 \centering
        \includegraphics[width=0.75\textwidth]{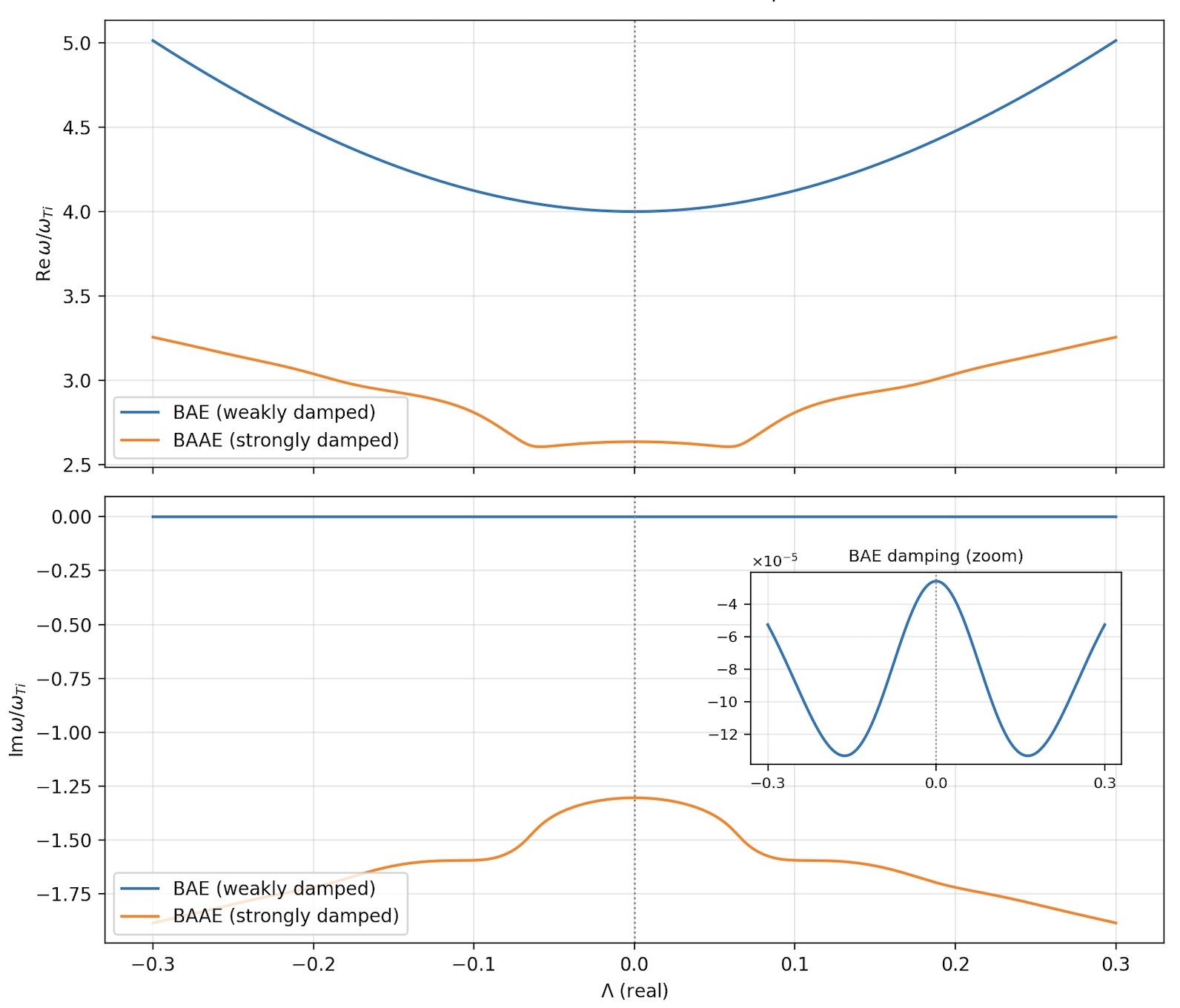}
 \caption{Continuous spectra for the kinetic BAE and BAAE branches as predicted from Eq. (\ref{eq:biglambdalow1}), showing the same qualitative features discussed in Fig. \ref{fig2}. Dimensionless parameters are
 $q = 2$, $\tau = 2$, $\beta_i = 0.01$, $\omega_{*ni}/\omega_{Ti} = 0.1$ and $\omega_{*Ti}/\omega_{Ti} = 0.2$. The top panel shows the stronger dependence of the BAE branch on $\Lambda$, parameterizing $qR_0 k_\parallel$; thus, 
 corresponding to a stronger excitation of BAE by EP with respect to BAAE \cite{chen17}. The bottom panel illustrates the strong BAAE ion Landau damping, while BAE damping is (exponentially) smaller, also due to different mode polarization.} 
\label{fig3}
\end{figure}

This limit was considered in Ref. \cite{zonca98} and is reproduced here for the readers' convenience.  Assuming well-circulating particles, $\omega_b = 
(m v_\parallel^2/2T_0)^{1/2} \omega_T$ with $\omega_T = (2 T_0/m)^{1/2}/(q R_0)$, $n \bar \omega_d = 0$ and 
$\theta_c = \sigma \vartheta_0$. Meanwhile, for circular geometry, $\omega_d = \hat \omega_d \sin \vartheta_0$ and $\hat \omega_d = - k_\vartheta s \vartheta_1 (v_\parallel^2+v_\perp^2/2)/(\bar \Omega R_0)$. 
Equation (\ref{eq:qnsolvelow}) then yields $\delta \bar \Phi_{\parallel 0}^{(0)} = 0$ and, assuming massless electrons, 
\begin{eqnarray}
\delta \hat K^{(1)} & = & \delta \hat K^{(1+)} e^{i \vartheta_0} + \delta \hat K^{(1-)}  e^{-i \vartheta_0} \; , \label{eq:dk1iont} \\
\delta \hat K^{(1\pm)} & =  & \frac{e}{m} \frac{Q \bar F_0}{\omega} \frac{1}{\pm \sigma \omega_b (1 \pm \Lambda) - \omega} 
\left( \omega \frac{|\bm \nabla r| }{\hat \kappa_\perp} \delta \bar \Phi_{\parallel 0}^{(1\pm)} \pm \frac{\hat \omega_d}{2i} \frac{|\bm \nabla r| }{\hat \kappa_\perp} \delta \bar \Psi_{0 }^{(0)} \right)\; , \nonumber
\end{eqnarray}
where we have omitted the thermal ion species subscript ``$i$" for brevity, and
\begin{eqnarray}
\delta \bar \Phi_{\parallel 0}^{(1)} & = & \delta \bar \Phi_{\parallel 0}^{(1+)} e^{i \vartheta_0} + \delta \bar \Phi_{\parallel 0}^{(1-)} e^{-i \vartheta_0} \; , \label{eq:phi1iont}  \\
 \delta \bar \Phi_{\parallel 0}^{(1\pm)}  & = & \mp i \frac{c T_{0i}}{e \bar B_0} \frac{N(\omega/\omega_{Ti}^{(\pm)})}{D(\omega/\omega_{Ti}^{(\pm)})} \frac{k_\vartheta}{\omega R_0} s \vartheta_1 \delta \bar \Psi_0^{(0)} \nonumber \; ;
\end{eqnarray}
consistent with the solution of Eq. (\ref{eq:qnsolvelow}). Here, $\omega_{Ti}^{(\pm)} = \omega_{Ti} (1 \pm \Lambda)$ and \cite{zonca96b,zonca98}
\begin{eqnarray}
N(x)&=&\left(1 - \frac{\omega_{*ni}}{\omega} \right) \left[ x + \left( 
1/2 + x^2 \right) Z(x) \right] -  \frac{\omega_{*Ti}}{\omega}
\left[ x \left( 1/2 + x^2 \right) + \left( 1/4 + x^4 \right) Z(x) \right]
\; ,  \nonumber \\
D(x)&=&\left(\frac{1}{x}\right) \left( 1 + \frac{1}{\tau} \right) 
+ \left(1 - \frac{\omega_{*ni}}{\omega} \right)  Z(x)  -  
\frac{\omega_{*Ti}}{\omega}
\left[ x + \left(  x^2 - 1/2 \right)  Z(x) \right]
\; , \label{eq:nd}
\end{eqnarray} 
with $Z(x)= \pi^{-1/2} \int_{-\infty}^\infty e^{-y^2}/(y-x) dy$ denoting the plasma dispersion function. 
Note that the present discussion keeps into account small but finite $\Lambda$ effects in the resonance condition, which were neglected in Ref. \cite{zonca98} near the accumulation point
of the continuous spectrum, but were kept into account in Refs. \cite{zonca14b,zonca10}, due to the important role it has in defining the sensitivity of mode excitation by EPs \cite{chen17} as shown above in Fig. \ref{fig2}.
When substituted back into the generalized inertia expression, Eq. (\ref{eq:Lambda2gen}), this yields \cite{zonca14b,zonca10}\footnote{For an extension of Eq. (\ref{eq:biglambdalow1}) to low mode numbers, interested readers are referred to the original works.}
\begin{eqnarray}
\Lambda^2 & = & \frac{\omega^2}{\omega_A^2}\left( 1 - \frac{\omega_{*pi}}{\omega} 
\right)  + q^2 \frac{\omega_{Ti}^2 }{2 \omega_A^2}
\left[ \left( 1 - \frac{\omega_{*ni}}{\omega} \right) \left( x^{(+)} F^{(+)} + x^{(-)} F^{(-)} \right) \right. \nonumber \\ & & \left . - 
\frac{\omega_{*Ti}}{\omega} \left(x^{(+)} G^{(+)} + x^{(-)} G^{(-)} \right)  - 
 \left( x^{(+)} N^{(+)} \frac{N^{(+)}}{D^{(+)}}  + x^{(-)} N^{(-)} \frac{N^{(-)}}{D^{(-)}}  \right)
\right] \; , \label{eq:biglambdalow1}
\end{eqnarray}
with $\omega_A = v_{A0}/(qR_0)$, $v_{A0}$ the on axis Alfv\'en speed, $x^{(\pm)} = \omega/\omega_{Ti}^{(\pm)}$, $N^{(\pm)} = N(x^{(\pm)})$ and likewise for the other functions, $D(x)$, $F(x)$ and $G(x)$, 
and the functions $F(x)$ and $G(x)$ defined as
\begin{eqnarray}
F(x) &=& x \left( x^2 + 3/2 \right) +  \left( x^4 + x^2 + 1/2 
\right) Z(x) \:\: ,
\label{eq:fg} \\
G(x) &=& x \left( x^4 + x^2 + 2 \right) +  \left( x^6 + x^4/2  
+ x^2 + 3/4 \right) Z(x)
\:\: . \nonumber \end{eqnarray}
Equation (\ref{eq:biglambdalow1}) can be solved for $\omega/\omega_{Ti}$ as a function of $\Lambda$, used as real variable parameterizing $qR_0 k_\parallel$. Results are shown in Fig. \ref{fig3} for 
BAE/BAAE branches with $q = 2$, $\tau = 2$, $\beta_i = 0.01$, $\omega_{*ni}/\omega_{Ti} = 0.1$ and $\omega_{*Ti}/\omega_{Ti} = 0.2$, illustrating the same qualitative behavior of Fig. \ref{fig2}.
The top panel confirms the different sensitivity of the BAE and BAAE branches on $\Lambda$ and, thereby, on the resonant excitation by EP in the present simplified gyrokinetic analysis of core plasma response. 
Meanwhile, the bottom panel shows the strong damping characterizing the BAAE branch, as opposed to BAE, which is weakly damped as shown in the inset.

At the next order, from Eq. (\ref{eq:fdklow2}) we obtain
\begin{eqnarray}
\delta \hat K^{(2)} & = & \delta \hat  K_0^{(2)} + \delta \hat K^{(2+)} e^{2i \vartheta_0} + \delta \hat K^{(2-)}  e^{-2i \vartheta_0} \; , \label{eq:dk2iont} \\
\delta \hat  K_0^{(2)}  & = & \frac{e}{m} \frac{Q \bar F_0}{\omega}  \left( -  \frac{|\bm \nabla r|}{\hat \kappa_\perp} \delta \bar \Phi_{\parallel 0 0}^{(2)}
\right)  + \frac{\hat \omega_d}{2 i \omega} \left( \delta \hat K^{(1-)} - \delta \hat K^{(1+)} \right)\; , \nonumber \\
\delta \hat K^{(2\pm)} & =  &  \frac{1}{\pm \sigma \omega_b (2 \pm \Lambda) - \omega} 
\left( \frac{e}{m} Q \bar F_0 \frac{|\bm \nabla r| }{\hat \kappa_\perp} \delta \bar \Phi_{\parallel 0}^{(2\pm)} \mp \frac{\hat \omega_d}{2i} \delta \hat K^{(1\pm)} \right)\; , \nonumber
\end{eqnarray}
where we have dropped the $\omega = \pm \Lambda \omega_{bi}$ Landau resonance, consistent with the present frequency ordering and well-circulating particle assumption. Equation (\ref{eq:phibar2}) then yields
\begin{eqnarray}
\delta \bar \Phi_{\parallel 0}^{(2)} & = & \delta \bar \Phi_{\parallel 0 0}^{(2)} + \delta \bar \Phi_{\parallel 0}^{(2+)} e^{2i \vartheta_0} + \delta \bar \Phi_{\parallel 0}^{(2-)} e^{-2i \vartheta_0} \; , \label{eq:phi2iont}  \\
\delta \bar \Phi_{\parallel 0 0}^{(2)} & = & -  \left( \frac{1}{\tau} + \frac{\omega_{*ni}}{\omega} \right)^{-1} \frac{\omega_A^2}{\omega^2} \Lambda^2 s^2 \vartheta_1^2 k_\vartheta^2 \bar \rho_{Li}^2 \delta \bar \Psi_0^{(0)}  \; ,  \nonumber \\
 \delta \bar \Phi_{\parallel 0}^{(2\pm)}  & = & s^2 \vartheta_1^2 k_\vartheta^2 \bar \rho_{Li}^2 \frac{q^2 (2 \pm \Lambda)}{2 D_{1/2}^{(\pm)}} \frac{\omega_{Ti}^2}{\omega^2} 
 \left[  \left( 1 - \frac{\omega_{*ni}}{\omega} \right) \left( F_{1/2}^{(\pm)} - F^{(\pm)} \right) \right.  \nonumber \\ & & \left.  - \frac{\omega_{*Ti}}{\omega}  \left( G_{1/2}^{(\pm)} - G^{(\pm)}  \right)  -  \frac{N^{(\pm)}}{D^{(\pm)}}  \left( N_{1/2}^{(\pm)} - N^{(\pm)}  \right)  \right] \delta \bar \Psi_0^{(0)} \nonumber \; ,
\end{eqnarray}
where $D_{1/2}^{(\pm)} = D (\omega/(2\pm \Lambda)\omega_{Ti})$ and likewise for the other functions, $N(x)$, $F(x)$ and $G(x)$, when a subscript $1/2$ is attached.
At third order, from Eq. (\ref{eq:fdklow3}), we finally have
\begin{eqnarray}
\delta \hat K^{(3)} & = & \delta \hat K^{(3+)} e^{i \vartheta_0} + \delta \hat K^{(3-)}  e^{-i \vartheta_0} + \delta \hat K^{(33+)} e^{3i \vartheta_0} + \delta \hat K^{(33-)}  e^{-3i \vartheta_0} \; , \label{eq:dk3iont} \\
\delta \hat K^{(3\pm)} & =  &  \frac{1}{\pm \sigma \omega_b (1 \pm \Lambda) - \omega} 
\left\{ \frac{e}{m} Q \bar F_0 \left[  \frac{|\bm \nabla r| }{\hat \kappa_\perp} \delta \bar \Phi_{\parallel 0}^{(3\pm)} -  \left(\frac{k_\vartheta^2 s^2 \vartheta_1^2 v_\perp^2}{4 \bar \Omega^2} \right) \left( \frac{|\bm \nabla r|}{\hat \kappa_\perp} \delta \bar \Phi_{\parallel 0}^{(1\pm)} \right. \right. \right. \nonumber \\
& & \left. \left. \left. \pm \frac{\hat \omega_d}{2i \omega} \frac{|\bm \nabla r|}{\hat \kappa_\perp}  \delta \bar \Psi_0^{(0)}  \right)  \right] \pm \frac{\hat \omega_d}{2i} \left( \delta \hat K^{(2\pm)} - \delta \hat K_0^{(2)}  \right) \right\}
 , \nonumber
\end{eqnarray}
where the explicit expressions of $\delta \hat K^{(33\pm)}$ are not given, since they are not needed for determining the wave dispersive properties at the order required here.
Equation (\ref{eq:phibar3}), thus, implies
\begin{eqnarray}
\delta \bar \Phi_{\parallel 0}^{(3)} & = & \delta \bar \Phi_{\parallel 0}^{(3+)} e^{i \vartheta_0} + \delta \bar \Phi_{\parallel 0}^{(3-)} e^{-i \vartheta_0} +  \delta \bar \Phi_{\parallel 0}^{(33+)} e^{3i \vartheta_0} + \delta \bar \Phi_{\parallel 0}^{(33-)} e^{-3i \vartheta_0} \; , \label{eq:phi3iont}  \\
\delta \bar \Phi_{\parallel 0}^{(3\pm)}  & = & - \sqrt{2} k_\vartheta \bar \rho_{Li} q  s \vartheta_1 \frac{\omega_{Ti}}{2 i \omega D^{(\pm)}} \left[ \left( 1 \pm \Lambda \right) \left(  N_{1/2}^{(\pm)} - N^{(\pm)} \right) \left(\pm \delta \bar \Phi^{(2 \pm)} \right)  
- N^{(\pm)}   \left(\pm \delta \bar \Phi^{(2)}_{\parallel 00} \right)  \right] \nonumber \\ & &
 - \sqrt{2} k_\vartheta \bar \rho_{Li} q  s \vartheta_1 \frac{\omega_{Ti}}{2 i \omega D^{(\pm)}} \left( k_\vartheta^2 \bar \rho_{Li}^2 s^2 \vartheta_1^2 \right)  \left(\pm \delta \bar \Psi^{(0)}_{0} \right) 
 \left[ \left( 1 - \frac{\omega_{*ni}}{\omega} \right) \left( V^{(\pm)} - \frac{N^{(\pm)}}{D^{(\pm)}} Z^{(\pm)} \right)  \right. \nonumber \\
 & & \left. - \frac{\omega_{*Ti}}{\omega}  \left( W^{(\pm)} - \frac{N^{(\pm)}}{D^{(\pm)}} (V^{(\pm)} - Z^{(\pm)}/2) \right)   \right]  + \sqrt{2} k_\vartheta \bar \rho_{Li} q 
  s \vartheta_1 \frac{\omega_{Ti}}{2 i \omega D^{(\pm)}} \left( k_\vartheta^2 \bar \rho_{Li}^2 s^2 \vartheta_1^2 \right) \nonumber \\
  & & \times \frac{q^2 \omega_{Ti}^2}{4 \omega^2} \left(\pm \delta \bar \Psi^{(0)}_{0} \right) \left\{  (2 \pm \Lambda) \left[ \left( 1 - \frac{\omega_{*ni}}{\omega} \right) \left( H_{1/2}^{(\pm)} - H^{(\pm)} 
  \right)- \frac{\omega_{*Ti}}{\omega}  \left(  I_{1/2}^{(\pm)} - I^{(\pm)}  \right)  \right] \right. \nonumber \\ & & \left. 
 - (2 \pm \Lambda) \frac{N^{(\pm)}}{D^{(\pm)}} \left[ \left( 1 - \frac{\omega_{*ni}}{\omega} \right) \left( F_{1/2}^{(\pm)} - F^{(\pm)} 
  \right) - \frac{\omega_{*Ti}}{\omega}  \left(  G_{1/2}^{(\pm)} - G^{(\pm)}  \right)  \right] \right. 
  \nonumber \\ & & \left. +  \left( 1 - \frac{\omega_{*ni}}{\omega} \right) \left( \frac{H^{(+)}  + H^{(-)}}{2}
  \right) - \frac{\omega_{*Ti}}{\omega}  \left( \frac{I^{(+)}  + I^{(-)}}{2}
  \right) -  \frac{N^{(\mp)}}{D^{(\mp)}}  \left[  \left( 1 - \frac{\omega_{*ni}}{\omega} \right)  \right. \right. \nonumber \\
 & & \left. \left.  \times    \left( \frac{F^{(+)}  + F^{(-)}}{2}
  \right) -   \frac{\omega_{*Ti}}{\omega}    \left( \frac{G^{(+)}  + G^{(-)}}{2}
  \right) \right]  \pm \frac{\Lambda}{1\pm \Lambda}  \left[  \left( 1 - \frac{\omega_{*ni}}{\omega} \right)  x^{(\pm)} H^{\prime(\pm)} \right.  \right. \nonumber \\
  & & \left. \left.  -   \frac{\omega_{*Ti}}{\omega} \right] x^{(\pm)} I^{\prime(\pm)}  
   \mp \frac{\Lambda}{1\pm \Lambda} \frac{N^{(\pm)}}{D^{(\pm)}}  \left[  \left( 1 - \frac{\omega_{*ni}}{\omega} \right) x^{(\pm)}  F^{\prime(\pm)}  -   \frac{\omega_{*Ti}}{\omega}  x^{(\pm)}  G^{\prime(\pm)}  \right]
   \right\} \; , \nonumber 
\end{eqnarray}
where
\begin{eqnarray}
V(x) & = & x + \left( x^{2} + 1 \right) Z(x) \:\:,  \label{eq:VWHI} \\
W(x) & = & x^{3} + 2 x + \left( x^{4} + (3/2) x^{2} + 3/2 \right) Z(x)
\:\: , \nonumber \\
H(x) & = & x^{5} + 2 x^{3} + 3 x + \left( x^{6} + (3/2) x^{4}
+ (3/2) x^{2} + 3/4 \right) Z(x) \:\: , \nonumber \\
I(x) & = & x^{7} + (3/2) x^{5} + (7/2) x^{3} + (27/4) x 
 + \left(
x^{8} + x^{6} + (9/4) x^{4} + 3 x^{2} + 15/8 \right) Z(x)
\:\: ; \nonumber 
\end{eqnarray}
the functions $F'(x)$, $G'(x)$, $H'(x)$ and $I'(x)$ denote the first derivatives of the respective functions defined above, which can be computed noting $Z'(x) = - 2 ( 1 + x Z(x) )$;
and, again, the explicit expressions of $ \delta \bar \Phi_{\parallel 0}^{(33\pm)}$ are not needed and, thus, are omitted for brevity. With these results,  noting Eq. (\ref{eq:gkvortKAW6}), the vorticity equation describing KAW propagation and absorption in the long wavelength limit, Eq. (\ref{eq:gkvortKAWlongapp}), yields \cite{zonca98}
\begin{eqnarray}
Q^2 (\omega) & = & \frac{\omega^2}{\omega_A^2} \left[ \frac{3}{4} \left( 1 - \frac{\omega_{*pi}}{\omega} - \frac{\omega_{*Ti}}{\omega} \right) + \frac{\Lambda^4 \omega_A^4/\omega^4}{(1/\tau+ \omega_{*ni}/\omega)} 
+ q^2 \frac{\omega_{Ti}}{\omega} S(\omega) \right] \label{eq:qq} \; , \\
S(\omega) \delta \bar \Psi_0^{(0)}  & = & \frac{1}{\left( k_\vartheta^2 \bar \rho_{Li}^2 s^2 \vartheta_1^2 \right)^2} \left\langle \frac{T_{0i}}{n_0 e} \frac{\hat \omega_{di}}{\omega_{Ti}} \frac{|s \vartheta_1|}{q^2}
\left[ \left( \frac{\delta \hat K^{(3-)} - \delta \hat K^{(3+)}}{2i} \right) \right. \right. \nonumber \\ & & \left. \left. 
- \frac{k_\vartheta^2 s^2 \vartheta_1^2 v_\perp^2}{4\bar \Omega^2} \left( \frac{\delta \hat K^{(1-)} - \delta \hat K^{(1+)}}{2i} \right) \right]_i \right\rangle_v \label{eq:qq} \; ; \label{eq:sfow}
\end{eqnarray}
with $\left\langle ... \right\rangle_v$ denoting velocity space integration. Substituting the relevant expression for the particle response derived in this Appendix above, 
and carrying out velocity space integrations, one can obtain
\begin{eqnarray}
S(\omega) & = &  \frac{q^2 \omega_{Ti}^2}{4 \omega^2} \left\{ (2 + \Lambda) \left[ \left( 1 - \frac{\omega_{*ni}}{\omega} \right) \left( L^{(+)} - L^{(+)}_{1/2} \right) - \frac{\omega_{*Ti}}{\omega} \left( M^{(+)} - M^{(+)}_{1/2} \right)  \right]
\right. \nonumber \\ & & + (2 - \Lambda) \left[ \left( 1 - \frac{\omega_{*ni}}{\omega} \right) \left( L^{(-)} - L^{(-)}_{1/2} \right) - \frac{\omega_{*Ti}}{\omega} \left( M^{(-)} - M^{(-)}_{1/2} \right)  \right]
\nonumber \\ & & - 2 (2 + \Lambda) \frac{N^{(+)}}{D^{(+)}} \left[ \left( 1 - \frac{\omega_{*ni}}{\omega} \right) \left( H^{(+)} - H^{(+)}_{1/2} \right) - \frac{\omega_{*Ti}}{\omega} \left( I^{(+)} - I^{(+)}_{1/2} \right)  \right] 
 \nonumber \\ & & - 2 (2 - \Lambda) \frac{N^{(-)}}{D^{(-)}} \left[ \left( 1 - \frac{\omega_{*ni}}{\omega} \right) \left( H^{(-)} - H^{(-)}_{1/2} \right) - \frac{\omega_{*Ti}}{\omega} \left( I^{(-)} - I^{(-)}_{1/2} \right)  \right] 
 \nonumber \\ & & + (2 + \Lambda) \frac{N^{(+)2}}{D^{(+)2}} \left[ \left( 1 - \frac{\omega_{*ni}}{\omega} \right) \left( F^{(+)} - F^{(+)}_{1/2} \right) - \frac{\omega_{*Ti}}{\omega} \left( G^{(+)} - G^{(+)}_{1/2} \right)  \right]  
 \nonumber \\ & & +  (2 - \Lambda) \frac{N^{(-)2}}{D^{(-)2}} \left[ \left( 1 - \frac{\omega_{*ni}}{\omega} \right) \left( F^{(-)} - F^{(-)}_{1/2} \right) - \frac{\omega_{*Ti}}{\omega} \left( G^{(-)} - G^{(-)}_{1/2} \right)  \right]   
 \nonumber \\ & & - \left[ \left( 1 - \frac{\omega_{*ni}}{\omega} \right) \left( L^{(+)} + L^{(-)} \right) - \frac{\omega_{*Ti}}{\omega} \left( M^{(+)} + M^{(-)} \right) \right] 
 \nonumber \\ & & + \left( \frac{N^{(+)}}{D^{(+)}} + \frac{N^{(-)}}{D^{(-)}}\right)  \left[ \left( 1 - \frac{\omega_{*ni}}{\omega} \right) \left( H^{(+)} + H^{(-)} \right) - \frac{\omega_{*Ti}}{\omega} \left( I^{(+)} + I^{(-)} \right) \right]
 \nonumber \\ & & - \frac{N^{(+)}N^{(-)}}{D^{(+)}D^{(-)}} \left[ \left( 1 - \frac{\omega_{*ni}}{\omega} \right) \left( F^{(+)} + F^{(-)} \right) - \frac{\omega_{*Ti}}{\omega} \left( G^{(+)} + G^{(-)} \right) \right]
 \nonumber \\ & & - \frac{\Lambda}{1+\Lambda} \left[ \left( 1 - \frac{\omega_{*ni}}{\omega} \right) x^{(+)} L^{\prime(+)}  - \frac{\omega_{*Ti}}{\omega}  x^{(+)} M^{\prime(+)} \right] 
 \nonumber \\ & & + \frac{\Lambda}{1-\Lambda} \left[ \left( 1 - \frac{\omega_{*ni}}{\omega} \right) x^{(-)} L^{\prime(-)} - \frac{\omega_{*Ti}}{\omega}  x^{(-)} M^{\prime(-)} \right]
 \nonumber \\ & & + \frac{2\Lambda}{1+\Lambda}  \frac{N^{(+)}}{D^{(+)}} \left[ \left( 1 - \frac{\omega_{*ni}}{\omega} \right) x^{(+)} H^{\prime(+)}  - \frac{\omega_{*Ti}}{\omega}  x^{(+)} I^{\prime(+)} \right] 
 \nonumber \\ & & - \frac{2\Lambda}{1-\Lambda}  \frac{N^{(-)}}{D^{(-)}} \left[ \left( 1 - \frac{\omega_{*ni}}{\omega} \right) x^{(-)} H^{\prime(-)} - \frac{\omega_{*Ti}}{\omega}  x^{(-)} I^{\prime(-)} \right]
 \nonumber \\ & & - \frac{\Lambda}{1+\Lambda}  \frac{N^{(+)2}}{D^{(+)2}} \left[ \left( 1 - \frac{\omega_{*ni}}{\omega} \right) x^{(+)} F^{\prime(+)}  - \frac{\omega_{*Ti}}{\omega}  x^{(+)} G^{\prime(+)} \right] 
 \nonumber \\ & & \left. + \frac{\Lambda}{1-\Lambda}  \frac{N^{(-)2}}{D^{(-)2}} \left[ \left( 1 - \frac{\omega_{*ni}}{\omega} \right) x^{(-)} F^{\prime(-)} - \frac{\omega_{*Ti}}{\omega}  x^{(-)} G^{\prime(-)} \right] \right\}
\nonumber \\ & & + \frac{q^2 \omega_{Ti}^2}{4 \omega^2} \left\{ \frac{(2  + \Lambda)}{D_{1/2}^{(+)}} \left[  \left( 1 - \frac{\omega_{*ni}}{\omega} \right) \left( F_{1/2}^{(+)} - F^{(+)} \right)  - \frac{\omega_{*Ti}}{\omega}  \left( G_{1/2}^{(+)} - G^{(+)}  \right)  \right. \right. \nonumber \\  & & \left.  -  \frac{N^{(+)}}{D^{(+)}}  \left( N_{1/2}^{(+)} - N^{(+)}  \right) \right]^2  + \frac{(2  - \Lambda)}{D_{1/2}^{(-)}} \left[  \left( 1 - \frac{\omega_{*ni}}{\omega} \right) \left( F_{1/2}^{(-)} - F^{(-)} \right)  \right. 
\nonumber \\  & & \left. \left. - \frac{\omega_{*Ti}}{\omega}  \left( G_{1/2}^{(-)} - G^{(-)}  \right)  -  \frac{N^{(-)}}{D^{(-)}}  \left( N_{1/2}^{(-)} - N^{(-)}  \right) 
\right]^2  \right\} \nonumber \\
& & +  \frac{\omega_{Ti}}{2\omega} \left[  \left( 1 - \frac{\omega_{*ni}}{\omega} \right)  \left( x^{(+)} T^{(+)} +  x^{(-)} T^{(-)} - 2 \frac{N^{(+)}}{D^{(+)}} x^{(+)} V^{(+)} 
 - 2 \frac{N^{(-)}}{D^{(-)}} x^{(-)} V^{(-)}  \right. \right. \nonumber \\
 & & \left. \left. +  \frac{N^{(+)2}}{D^{(+)2}} x^{(+)} Z^{(+)}  +  \frac{N^{(-)2}}{D^{(-)2}} x^{(-)} Z^{(-)}  \right) -   \frac{\omega_{*Ti}}{\omega}   \left( x^{(+)} U^{(+)} +  x^{(-)} U^{(-)} 
 \right. \right. \nonumber \\ & &  - 2 \frac{N^{(+)}}{D^{(+)}} x^{(+)} W^{(+)} - 2 \frac{N^{(-)}}{D^{(-)}} x^{(-)} W^{(-)} + \frac{N^{(+)2}}{D^{(+)2}} x^{(+)} (V^{(+)}- Z^{(+)}/2) 
 \nonumber \\ & & \left. \left. + \frac{N^{(-)2}}{D^{(-)2}} x^{(-)} (V^{(-)}- Z^{(-)}/2) \right) \right] + \frac{\Lambda^2\omega_A^2}{q^2 \omega \omega_{Ti}} \frac{\left[ \Lambda^2 \omega_A^2/\omega^2 
 - \left( 1 - \omega_{*pi}/\omega \right)\right]}{\left( 1/\tau + \omega_{*ni}/\omega\right)} \; ; \label{eq:sfow2}
\end{eqnarray}
where
\begin{eqnarray}
T(x) & = &  x^{3} + (5/2) x + \left( x^{4} + 2 x^{2} + (3/2) \right) Z(x)
\:\:, \label{eq:TULM} \\ 
U(x) & = & x^{5} + 3 x^{3} + (13/2) x + \left( x^{6} + (5/2) x^{4} +
(9/2) x^{2} + (15/4) \right) Z(x) 
\:\:, \nonumber \\ 
L(x) & = & x^{7} + (5/2) x^{5} + (19/4) x^{3} + (63/8) x + \left(
x^{8} + 2 x^{6} + 3 x^{4} + 3 x^{2} + 3/2 \right) Z(x)
\:\:, \nonumber \\
M(x) & = & x^{9} + 2 x^{7} + (11/2) x^{5} + (25/2) x^{3} +(201/8) x
+ \left( x^{10} + (3/2) x^{8} +  4 x^{6} + (15/2) x^{4} \right. \nonumber \\
& & \left. + 9 x^{2} 
+ 21/4 \right) Z(x)
\:\: . \nonumber
\end{eqnarray}

\funding{This work has been carried out within the framework of the EUROfusion Consortium, funded by the
European Union via the Euratom Research and Training Programme (Grant Agreement No. 101052200-
EUROfusion). Views and opinions expressed are, however, those of the author(s) only and do not necessarily 
reflect those of the European Union or the European Commission. Neither the European Union
nor the European Commission can be held responsible for them. This work 
was also supported by the MMNLP project CSN4 of INFN, Italy, the
National Key R\&D Program of China under Grant No. 2024YFE03170000,
the Strategic Priority Research Program of Chinese Academy of Sciences under Grant No.
XDB0790201, the National Science Foundation of China under Grant No. 12275236, the
CNNC Basic Research Program (No. CNNC-JCKY-202640), and the International Partnership Program of Chinese Academy of Sciences under Grant
No. 145GJHZ2025017BS.}

\roles{Fulvio Zonca: Methodology (lead); Validation (lead); Writing – original
draft (lead); Writing – review \& editing (lead). Liu Chen: Methodology (lead); Writing – review \& editing (equal). Matteo V. Falessi: Writing – review \& editing
(equal). Ruirui Ma: Writing – review \& editing (equal). Zhiyong Qiu: Writing – review \& editing (equal). Guangyu Wei: Writing – review \& editing (equal).}

\data{The data cannot be made publicly available upon publication because no suitable repository exists for
hosting data in this field of study. The data that support the findings of this study are available upon
reasonable request from the authors.}


\bibliographystyle{iopart-num} 
\bibliography{bibliography}

\end{document}